\documentclass[draft,grl]{agutex}
\usepackage[]{graphicx} 
\usepackage{overpic}
\usepackage{setspace} 
\usepackage{epstopdf}
\usepackage[margin=0cm,font={normalsize}, labelfont={bf,large}]{caption}
\usepackage{enumitem}

\usepackage{color}

\begin{document}

\authorrunninghead{MOSELEY ET AL.}

\titlerunninghead{CONVECTIVE EXTREMES}

\authoraddr{Corresponding author: Christopher Moseley,
Max Planck Institute for Meteorology
(christopher.moseley@mpimet.mpg.de)}

\title{Intensification of convective extremes driven by cloud-cloud interaction}


\authors{Christopher Moseley \altaffilmark{1}, Cathy Hohenegger \altaffilmark{1}, Peter Berg\altaffilmark{2}, Jan O. Haerter\altaffilmark{3}}

\altaffiltext{1}{Max Planck Institute for Meteorology, Hamburg, Germany and
Helmholtz Zentrum Geesthacht, Climate Service Center, Hamburg, Germany}
\altaffiltext{2}{Hydrology Research Unit, Swedish Meteorological and Hydrological Institute, Norrk\"oping, Sweden.}
\altaffiltext{3}{Niels Bohr Institute, Copenhagen, Denmark.}

\pagebreak
\begin{abstract} 
{\bf 
In a changing climate, a key role may be played by the response of convective-type cloud and precipitation to temperature changes \citep{OGorman2012,kendon2014heavier,mauritsen2015missing,tan2015increases}.
Yet, it is unclear if precipitation intensities will increase mainly due to modified thermodynamic forcing or due to stronger convective dynamics \citep{westra2014future}. 
Here we perform large eddy simulations (LES) of the convective dynamics by imposing an idealized diurnal cycle of surface temperature. 
In gradual self-organization, convective events produce highest intensities late in the day.
Tracking rain cells throughout their life cycles, we find that interacting events respond strongly to changes in boundary conditions.
Conversely, events without interaction remain unaffected.
Increased surface temperature indeed leads to more interaction and higher precipitation extremes. 
However, a similar intensification occurs when leaving temperature unchanged but simply granting more time for self-organization.
Our study implies that the convective field as a whole acquires a memory of past precipitation and inter-cloud dynamics, driving extremes.
Our results implicate that the dynamical interaction between convective clouds must be incorporated in global climate models to describe convective extremes and the diurnal cycle more realistically.
}
\end{abstract}

\begin{article}
\section{Introduction}\label{sec:intro}
Observations show strong increases in extreme convective precipitation intensity beyond thermodynamic expectations as temperatures rise \citep{berg2013strong, Lenderink:2008}. 
An increase of approximately $7$ $\%/K$ would be expected by the thermodynamic Clausius-Clapeyron relation for saturation vapor pressure.
Hence, dynamics may also contribute to the increase.
Modeling studies at convection resolving scales have made attempts at identifying processes that drive such strong increases of convective intensities. 
Some indeed show strong \citep{Singleton:2012} or abrupt \citep{meredith2015crucial} scaling.
Others highlight the potential sensitivity of results to model details \citep{attema2014,singh2014}, or fail to see the scaling altogether \citep{Muller:2011}.

Convective, in contrast to stratiform events, are characterized by buoyantly unstable atmospheric conditions, which can be induced by heating of moist surface air. 
As buoyant moist parcels are lifted and cooled, release of latent heat from condensation leads to further buoyancy, thereby driving the updrafts.
After onset of precipitation, its reevaporation and melting induces cooling --- counteracting further precipitation --- and generates cold pools, i.e. areas of relatively low temperature \citep{tompkins2001organizationCold,boing2012influence,schlemmer2014}.  
Cold pools have been stated to spread outward from precipitating cells as density currents, and were associated with the triggering of additional convection when they collide \citep{tompkins2001organizationCold, schlemmer2014}.
Through these processes, organization of the atmosphere may occur as a combination of thermodynamic and mechanical processes \citep{torri2015mechanisms}. These organising mechanisms, alongside with temporally varying boundary conditions as e.g. imposed by a diurnal cycle, imply that a convective system cannot in general be treated as being in a state of ``quasi-equilibrium'' \citep{yano2012}, but rather in a transitional state towards convective organization.

We simulate a single diurnal cycle of convection in a set of idealized LES numerical experiments.  
The state-of-the-art LES \citep{stevens2005} is run at 200~m horizontal resolution subject to temporally varying surface temperature and solar irradiation (Fig.~\ref{fig:sounding}).
We carry out three distinct simulations:
A control simulation, termed ``CTR'', with average surface temperature set to 23~$^{\circ}$C, is compared to a temperature-increased simulation, termed ``P2K'', where the surface temperature timeseries is offset by $+2$~K, as well as a simulation where all settings of ``CTR'' were maintained, but time was stretched so that a longer model day of 48~h resulted (termed ``LD'').
The initial conditions were provided by 
convectively unstable temperature profiles derived from radiosondes at a mid-latitude station. 
Lateral fluxes were initialized as zero, i.e. there was no horizontal wind (Details: Methods).

\begin{figure*}[h]
\begin{center}
\begin{overpic}[width=5cm,angle=-90,trim= 0cm 0pt 0pt 0pt,clip]{dummy.pdf}

\put(-62.9,40){\includegraphics[width=8.05cm,trim= 0cm 1.4cm 0cm 0cm,clip]{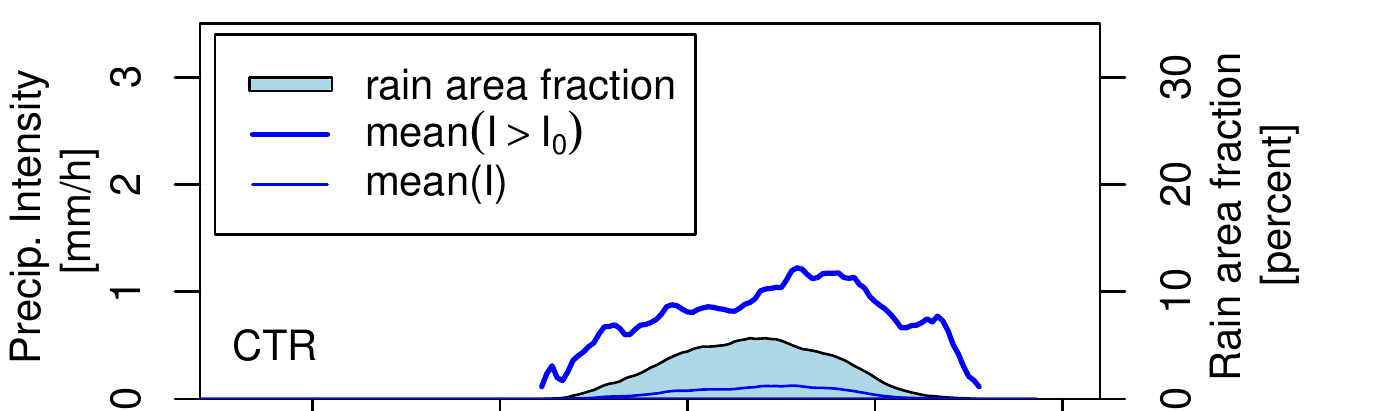}}
\put(-67.3,20){\includegraphics[width=8.35cm,trim= 0cm 1.4cm 0cm 0cm,clip]{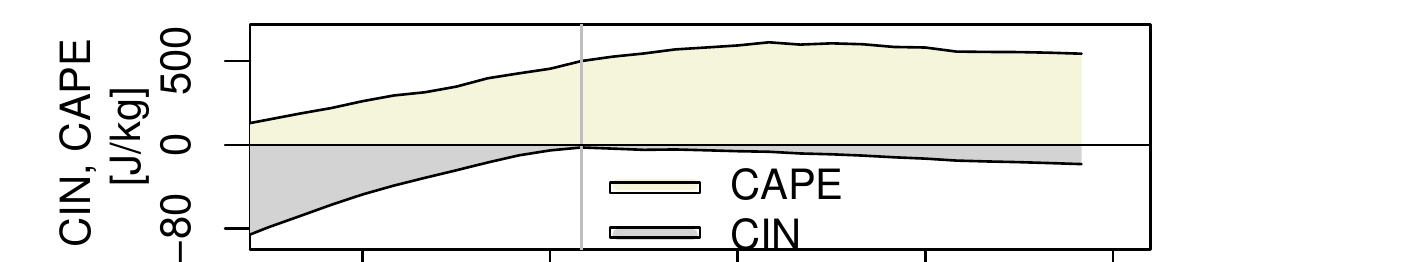}}

\put(-62.9,-13){\includegraphics[width=8.05cm,trim= 0cm 1.4cm 0cm 0cm,clip]{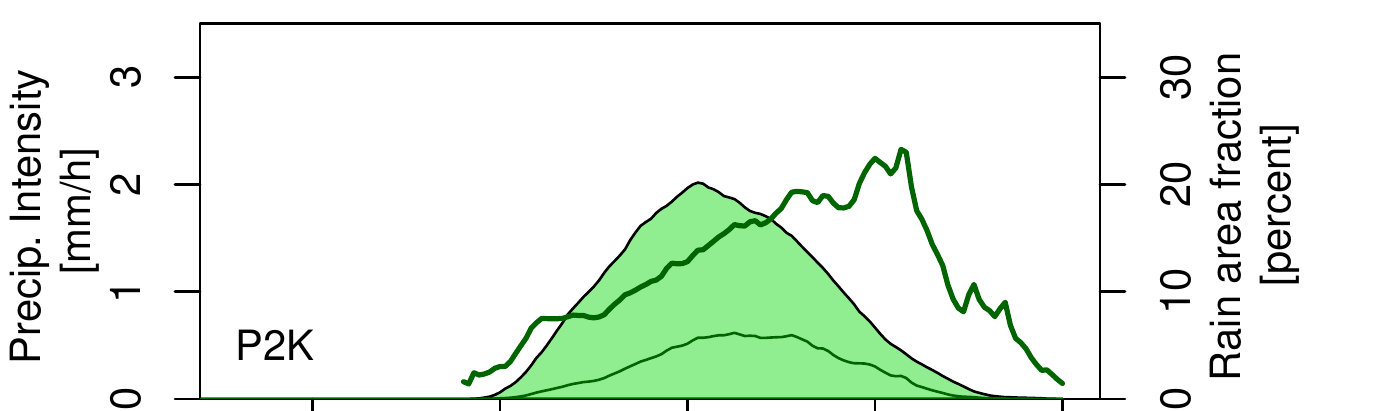}}
\put(-67.3,-33){\includegraphics[width=8.35cm,trim= 0cm 1.4cm 0cm 0cm,clip]{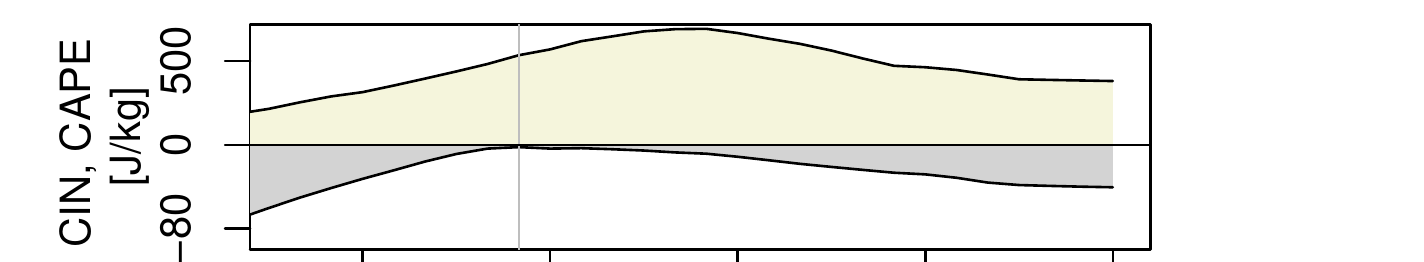}}

\put(-62.9,-66){\includegraphics[width=8.05cm,trim= 0cm 1.4cm 0cm 0cm,clip]{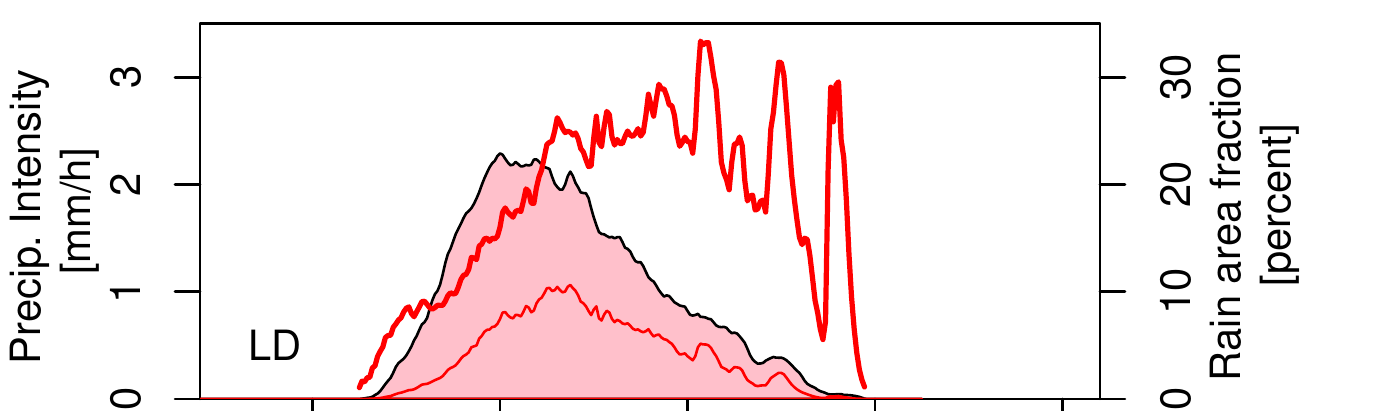}}
\put(-67.3,-97){\includegraphics[width=8.35cm,trim= 0cm 0pt 0cm 0cm,clip]{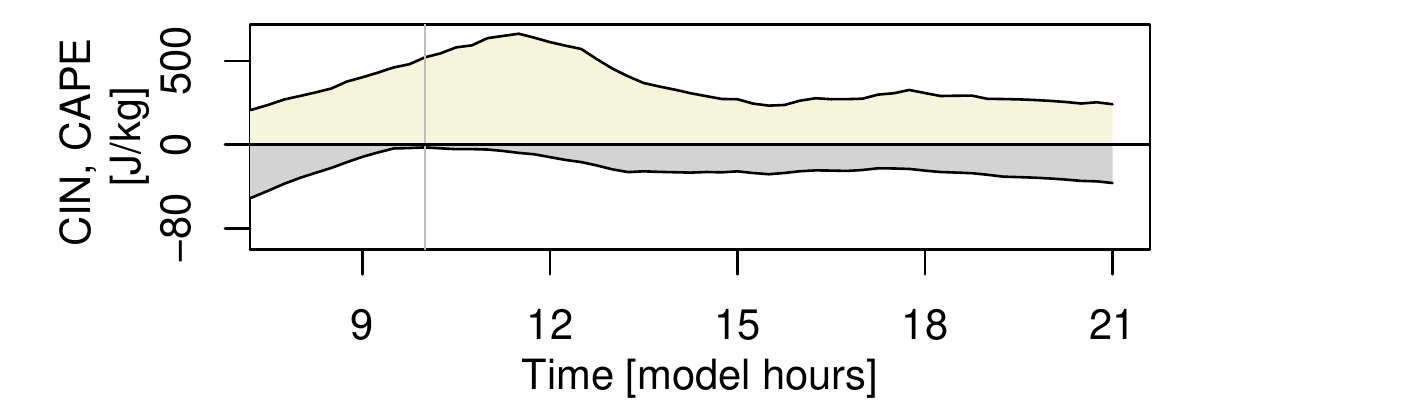}}

\put(-70,65){{\bf a}}
\put(-70,10){{\bf b}}
\put(-70,-42){{\bf c}}

\end{overpic}
\vspace{200pt}
\caption{\small{\bf Diurnal cycle of precipitation and potential energy.}
{\bf a}, Control simulation (CTR) showing area affected by rain, i.e. areas with $I>I_0$ where $I_0=.1\;mm\;h^{-1}$ (light blue, shaded, right axis) as well as domain mean precipitation (thin blue line) and conditional mean (bold blue line).
Domain mean precipitation intensity was multiplied by a factor two to enhance visibility.
Shaded beige and gray areas indicate convective available potential energy (CAPE) and convective inhibition energy (CIN).
Vertical gray line marks the time of minimal inhibition energy.
{\bf b}, Similar to (a) but for the temperature-increased (P2K) simulation.
{\bf c}, Similar to (a) but for the longer day (LD) simulation.
In (c), the model day is $48\;h$ and the horizontal axis was rescaled by a factor of $1/2$ to match panels (a) and (b).
}
\end{center}
\label{fig:summary_diurnal_cycle}
\end{figure*}

Surface fluxes are dominated by latent heat flux in all simulations. Both sensible and latent heat flux set in in the early morning, and reach their maximum shortly after noon (Fig. \ref{fig:surfacefluxes}).
We define ``precipitation yield'' as the domain average of precipitation, whereas ``precipitation intensity'' is measured only over the area where precipitation occurs. 
In all simulations, precipitation sets in before the time of maximum surface flux and decays during the second half of the model day (Fig. \ref{fig:summary_diurnal_cycle}).
Convective available potential energy (CAPE) and convective inhibition (CIN) are classical measures for parameterization of convective precipitation in large-scale models \citep{arakawa2004cumulus}. 
We therefore now first examine domain averages, roughly corresponding to GCM gridbox resolution.
When CIN is overcome, convection is triggered, allowing energy stored as CAPE to be released as latent heat. 
Precipitation onset generally coincides with weak, but nonzero, CIN and increasing CAPE. 
As precipitation continues, CAPE peaks and begins to decline. 
Common to all three simulations is that precipitation intensity peaks later than precipitation yield, hinting at processes that focus intensities.

Apart from these commonalities, the three experiments differ in important ways:
CTR produces overall small precipitation yield (Fig. \ref{fig:summary_diurnal_cycle}a).
CAPE accordingly only slightly decreases after the maximum is reached, i.e. only a fraction of CAPE is ``consumed'' by conversion to latent heat.
In P2K (Fig. \ref{fig:summary_diurnal_cycle}b), precipitation yield is substantially increased and most, but not all, of CAPE is ``consumed''. Notably, this is achieved despite overall stronger CIN compared to CTR.
LD produces the most intense precipitation which affects over 20 percent of the domain (Fig. \ref{fig:summary_diurnal_cycle}c). 
Even though CAPE is reduced considerably during the second half of the model period, precipitation intensity remains high.

\begin{figure}[t]
\begin{center}
\begin{overpic}[width=5cm,angle=-90,trim= 0cm 0pt 0pt 0pt,clip]{dummy.pdf}
\put(0,-125){\includegraphics[height=2.5cm,trim= 0cm 0cm 0cm 0cm,clip]{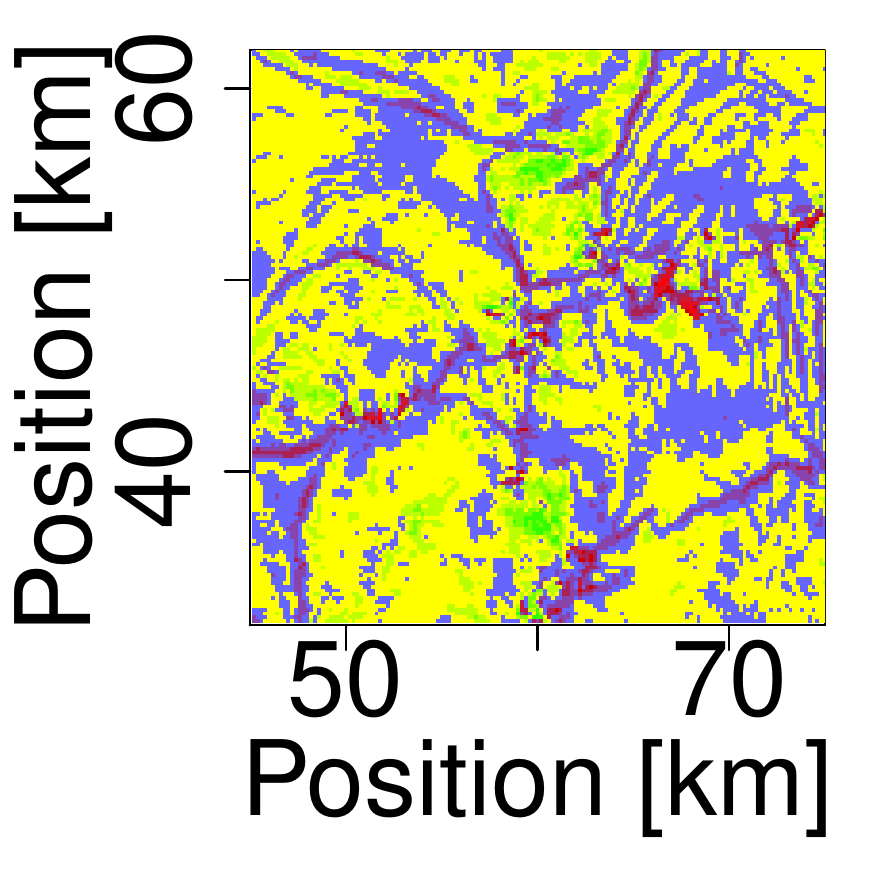}}
\put(34,-125){\includegraphics[height=2.5cm,trim= 2cm 0cm 0cm 0cm,clip]{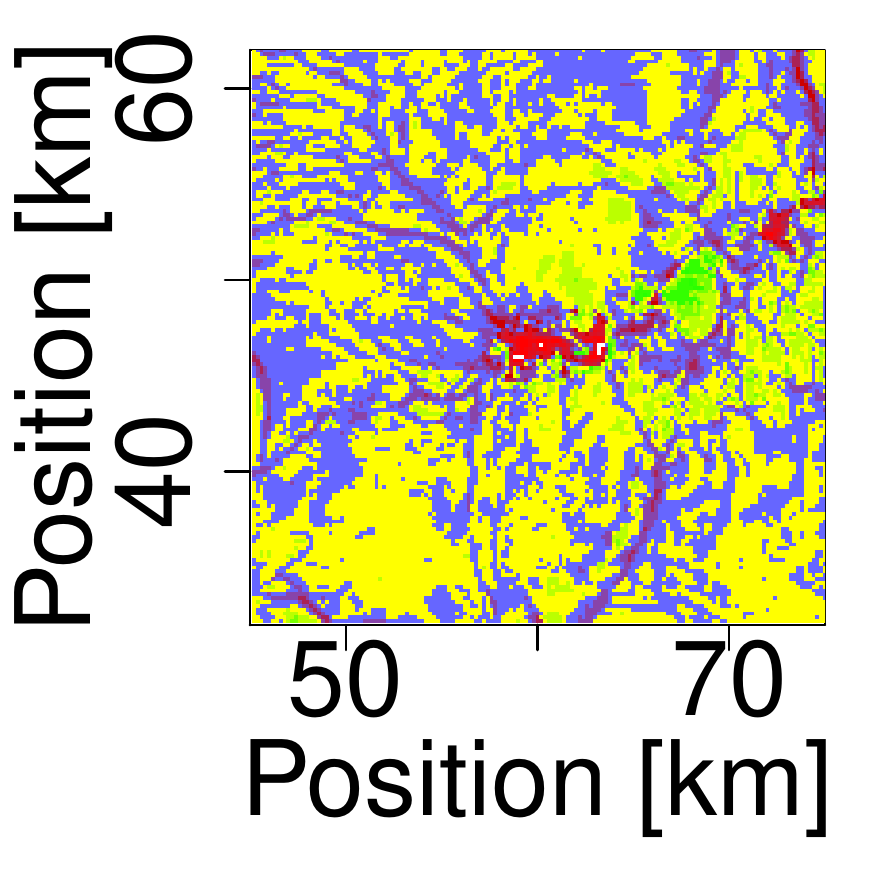}}
\put(60,-125){\includegraphics[  height=2.5cm,trim= 2cm 0cm 0cm 0cm,clip]{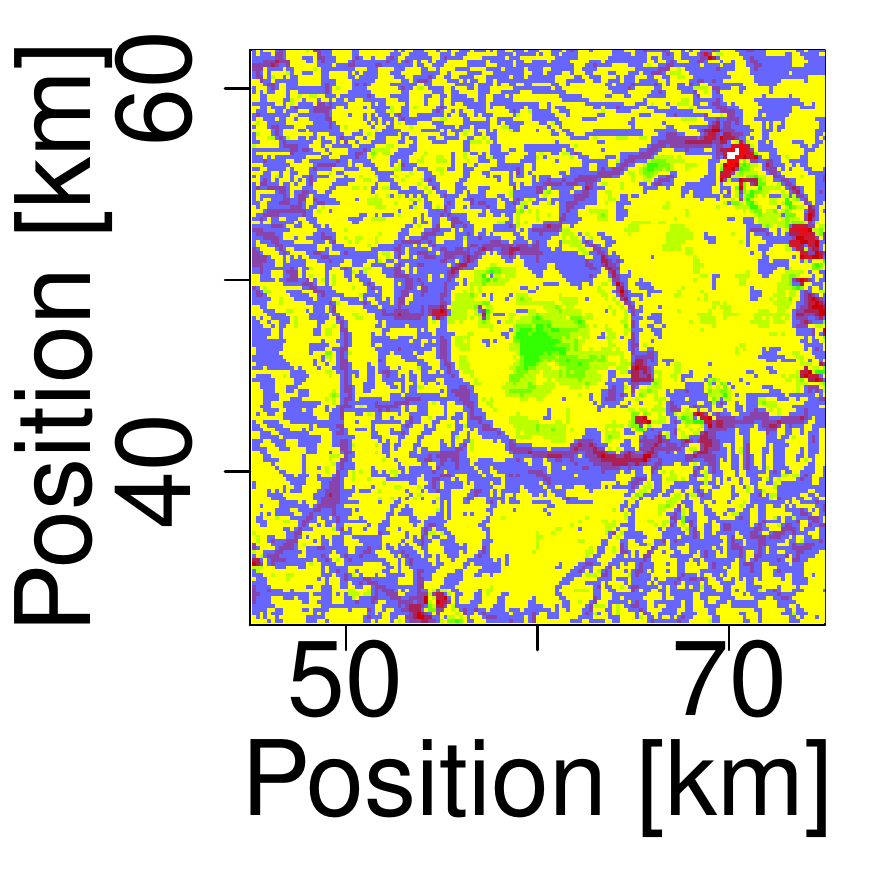}}

\put(0,-87){\includegraphics[height=1.8cm,trim= 0cm 2.45cm 0cm 0cm,clip]{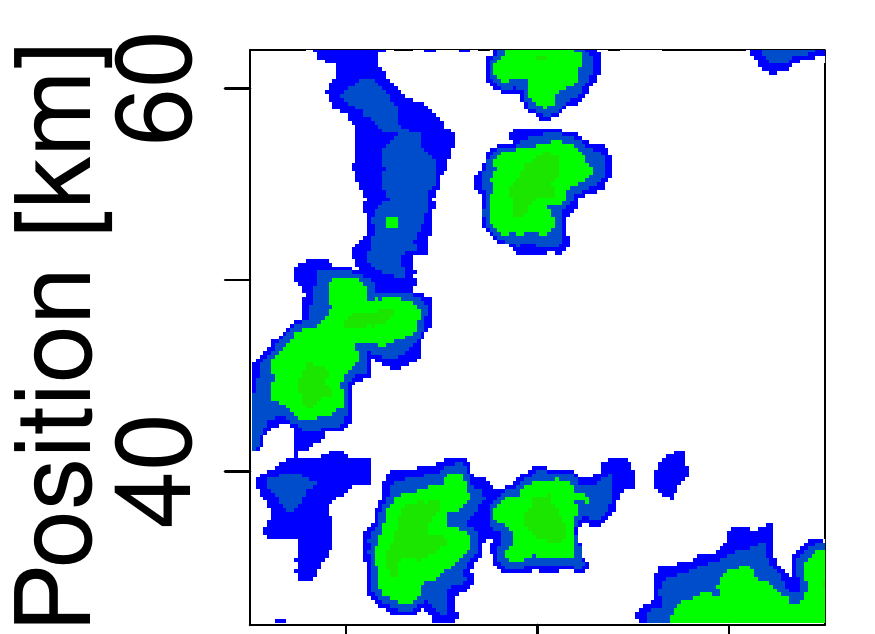}}
\put(34,-87){\includegraphics[height=1.8cm,trim= 2cm 2.45cm 0cm 0cm,clip]{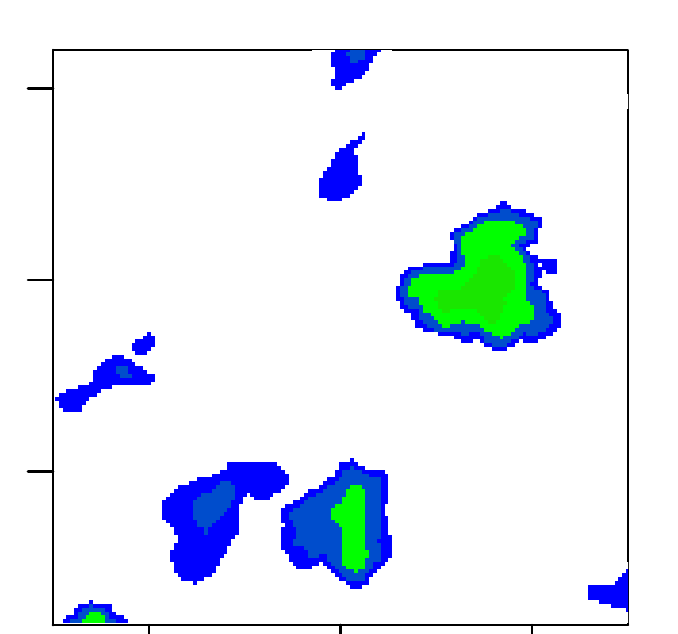}}
\put(60,-87){\includegraphics[height=1.8cm,trim= 2cm 2.45cm 0cm 0cm,clip]{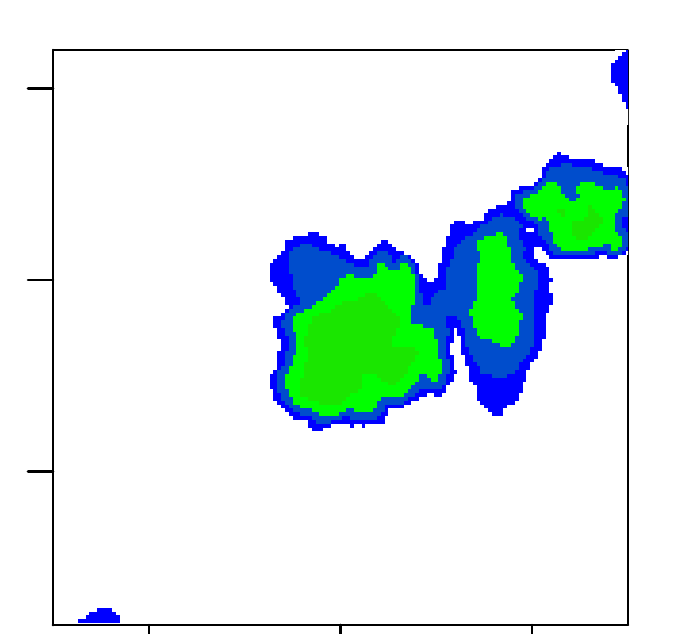}}

\put(  86,-98){\includegraphics[height=3cm,trim= 15cm 0pt 0cm 1.5cm,clip]{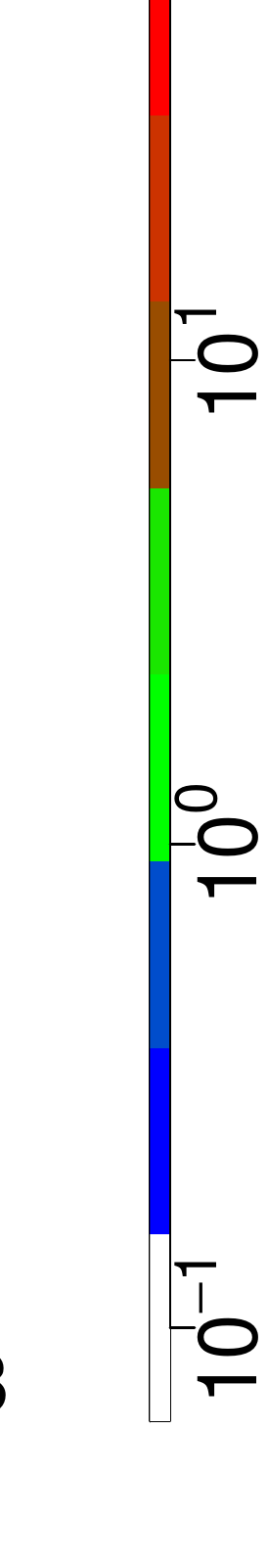}}

\put(-63,-93){\includegraphics[width=4.5cm,trim= 0cm 2cm 0cm 0cm,clip]{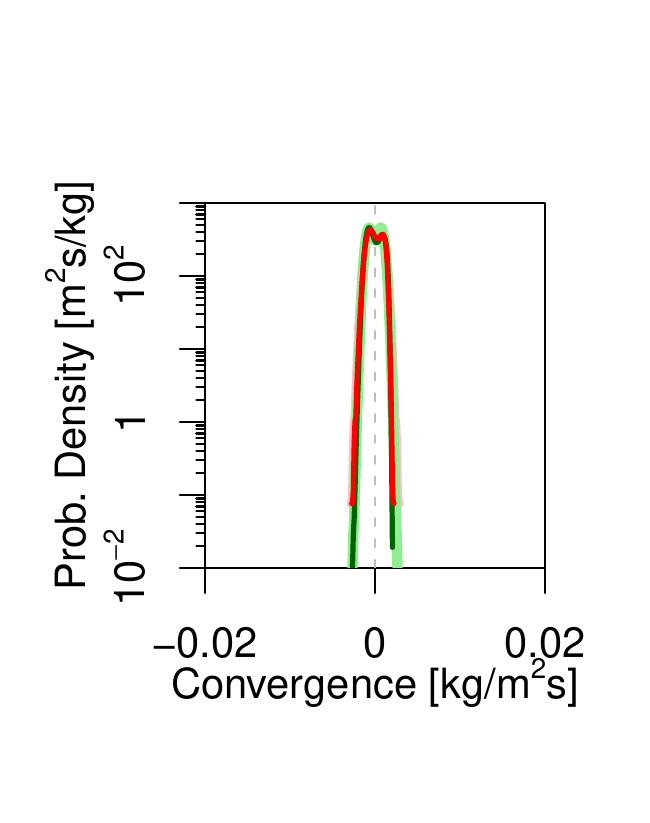}}
\put(-63,-153){\includegraphics[width=4.5cm,trim= 0cm 0pt 0cm 0cm,clip]{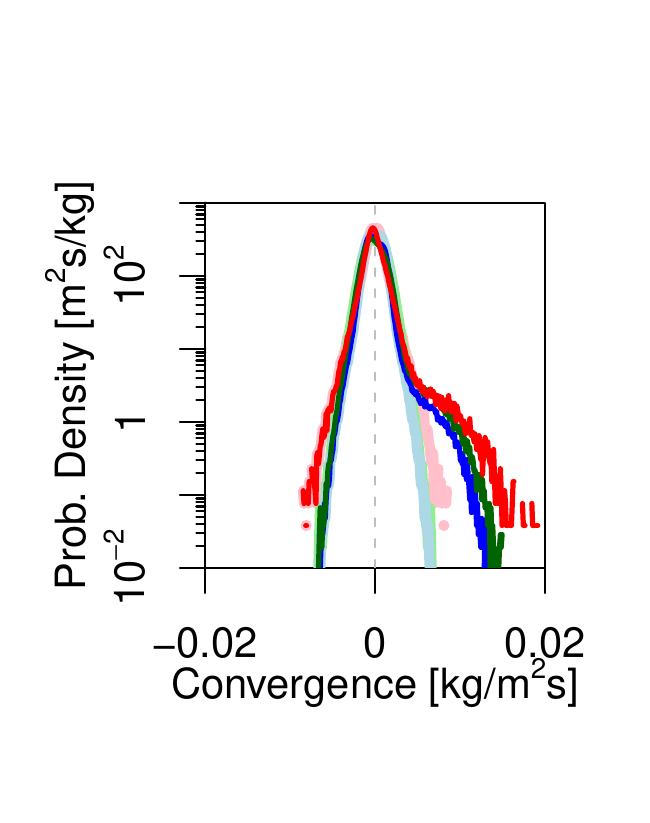}}

\put(-23,-57){{\bf \small $t<t_0$}}
\put(-23,-98){{\bf \small $t>t_0$}}
\put(-43,-57){(i)}
\put(-43,-98){(ii)}

\put( 102,-91){\includegraphics[width=4.7cm,trim= 0cm 1.4cm 0cm 0cm,clip]{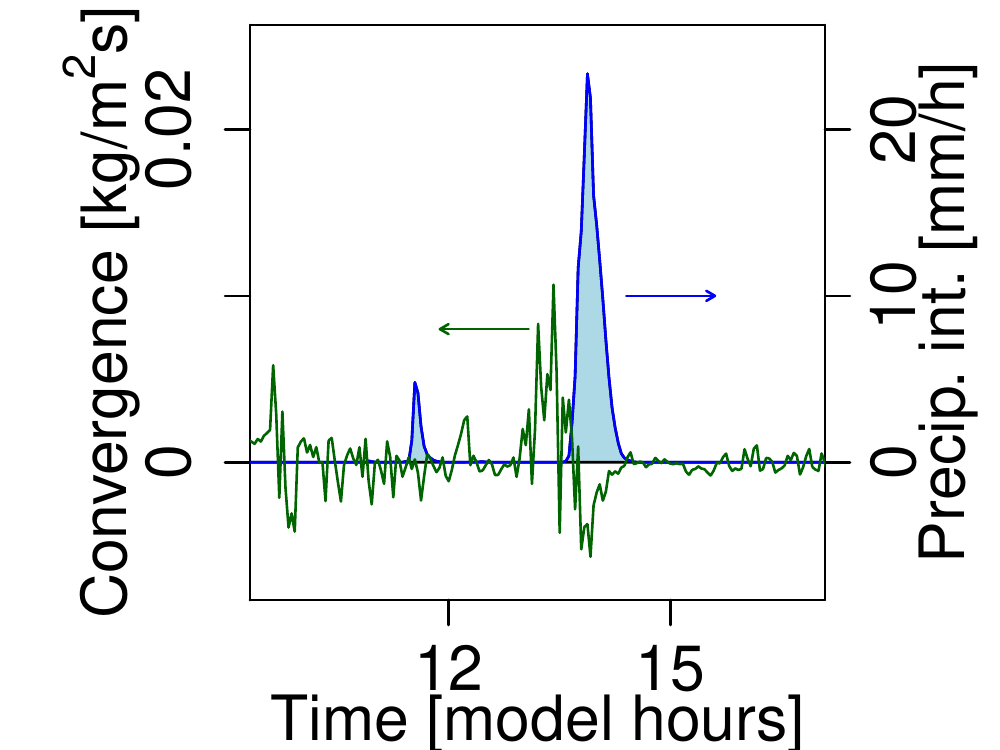}}
\put( 106.5,-112){\includegraphics[width=4.3cm,trim= 0cm 1.4cm 0cm 0cm,clip]{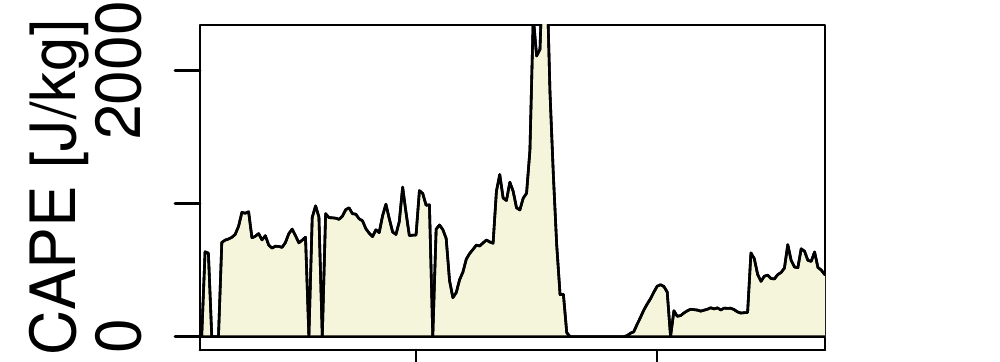}}
\put( 106.5,-141){\includegraphics[width=4.3cm,trim= 0cm 0pt 0cm 0cm,clip]{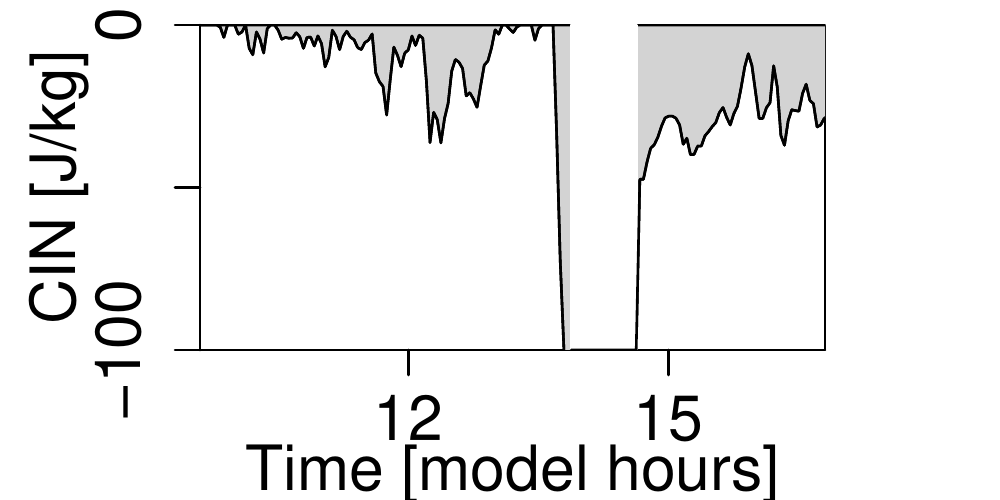}}

\put( -70,-45){\includegraphics[height=7cm,trim= 0cm 1.3cm 9cm 1.3cm,clip]{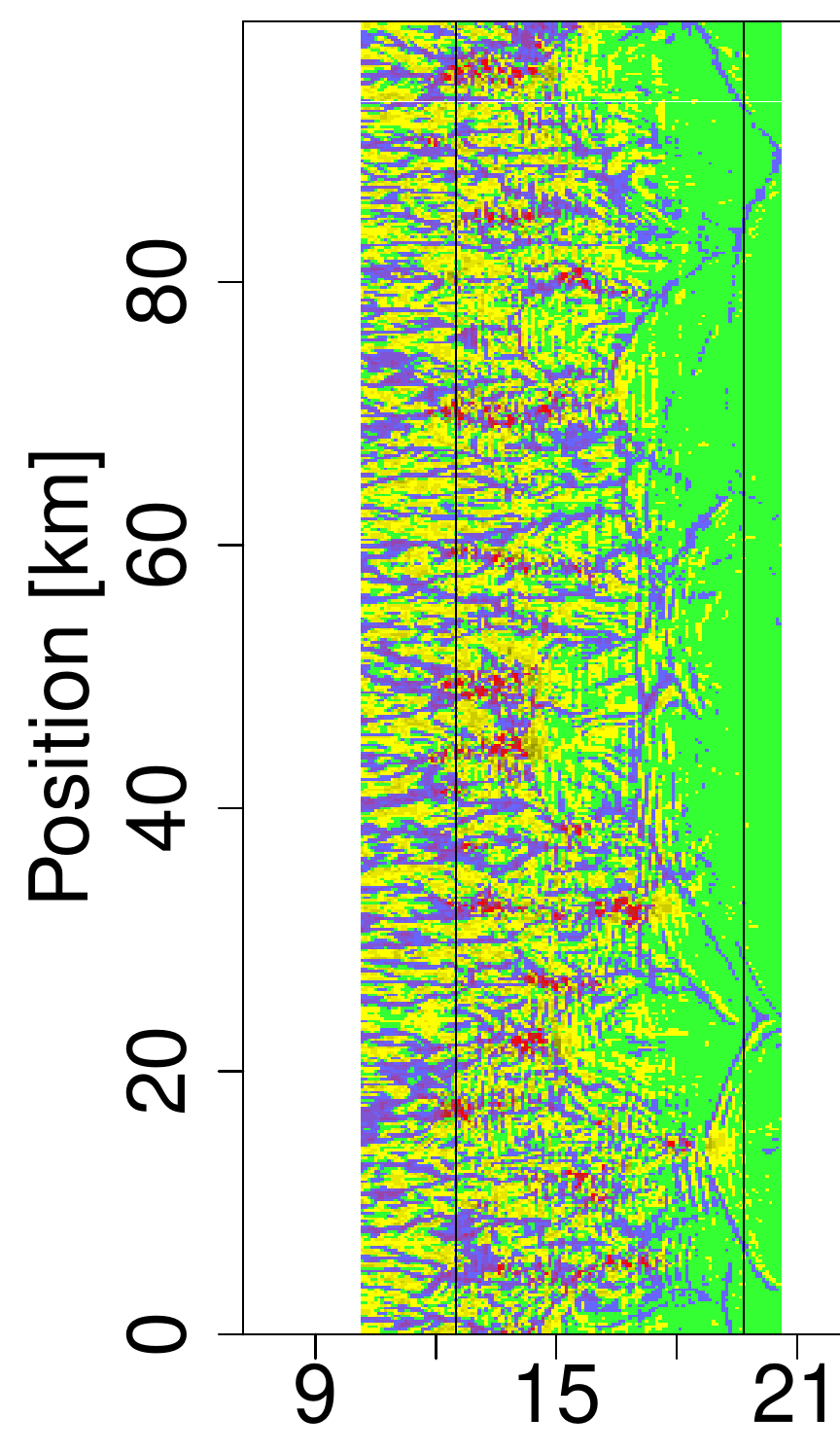}}
\put(   -10,-45){\includegraphics[height=7cm,trim= 2.5cm 1.3cm 9cm 1.3cm,clip]{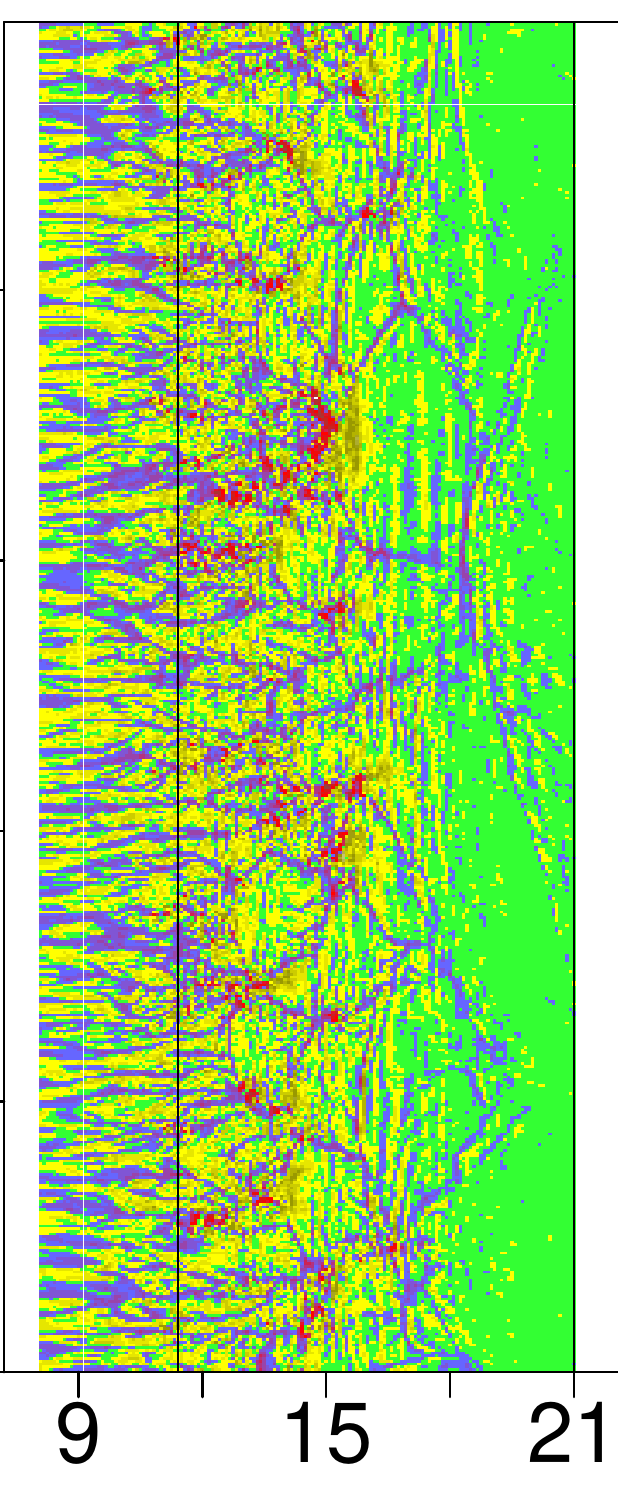}}
\put(  30,-45){\includegraphics[height=7cm,trim= 5.0cm 1.3cm 2cm 1.3cm,clip]{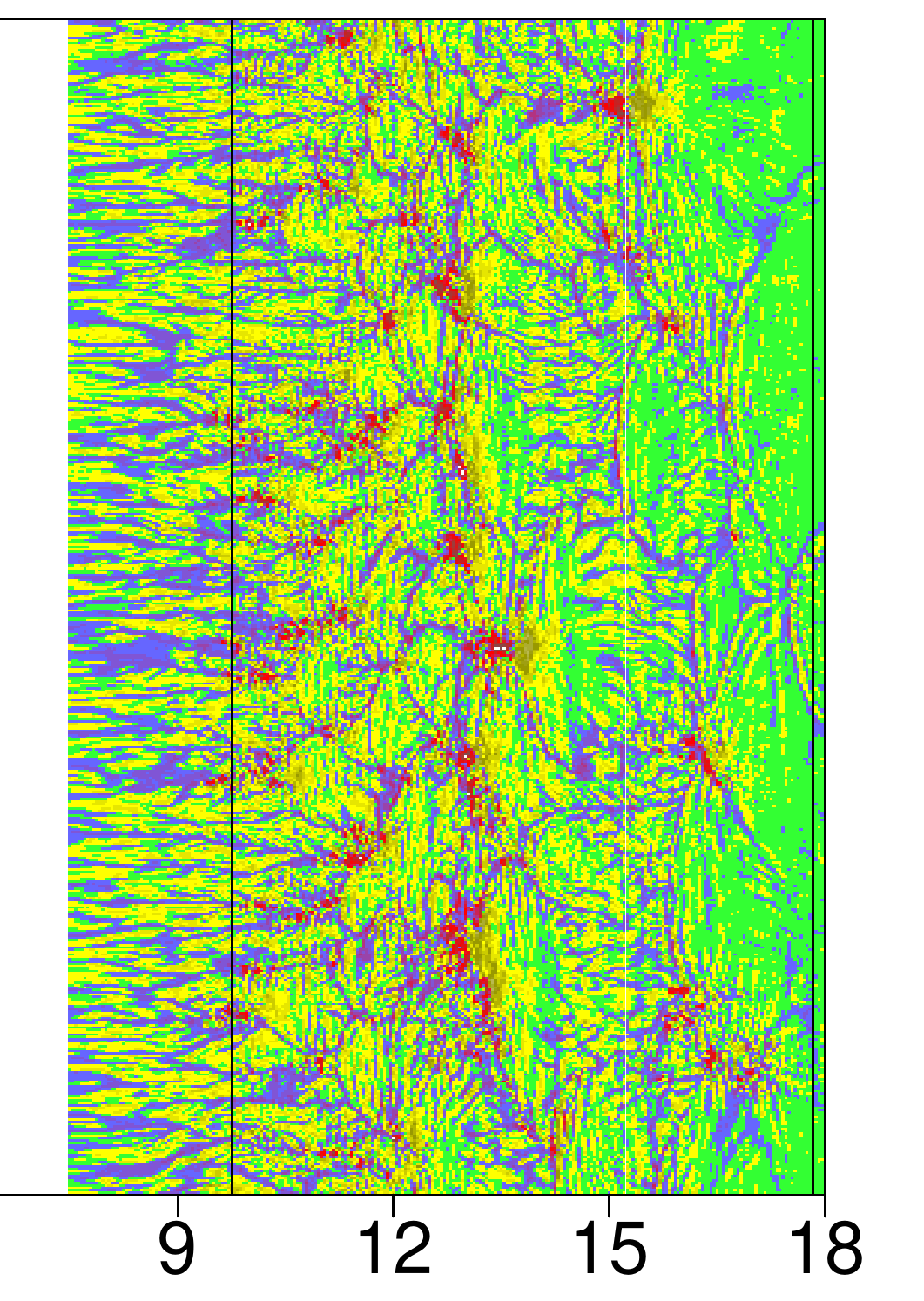}}
\put(  87,-30){\includegraphics[height=5.5cm,trim= 15.0cm 1.3cm 0cm 1.3cm,clip]{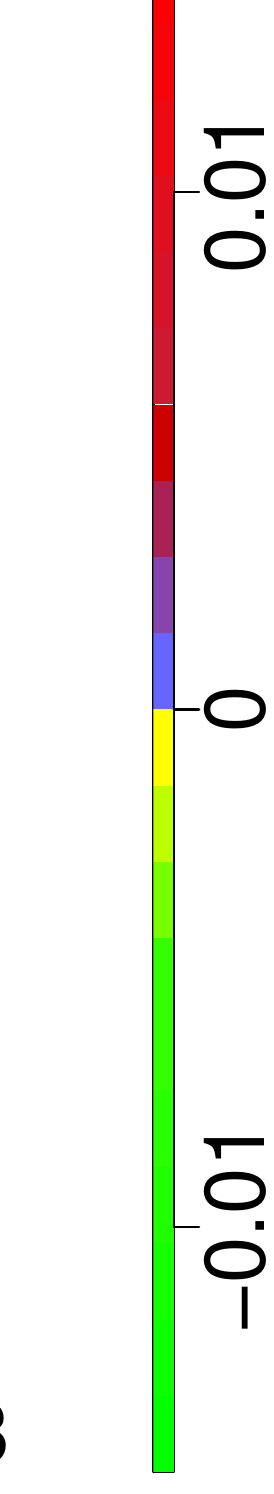}}

\put( 100,-45){\includegraphics[  height=7cm,trim= 5cm 1.3cm 2cm 1.3cm,clip]{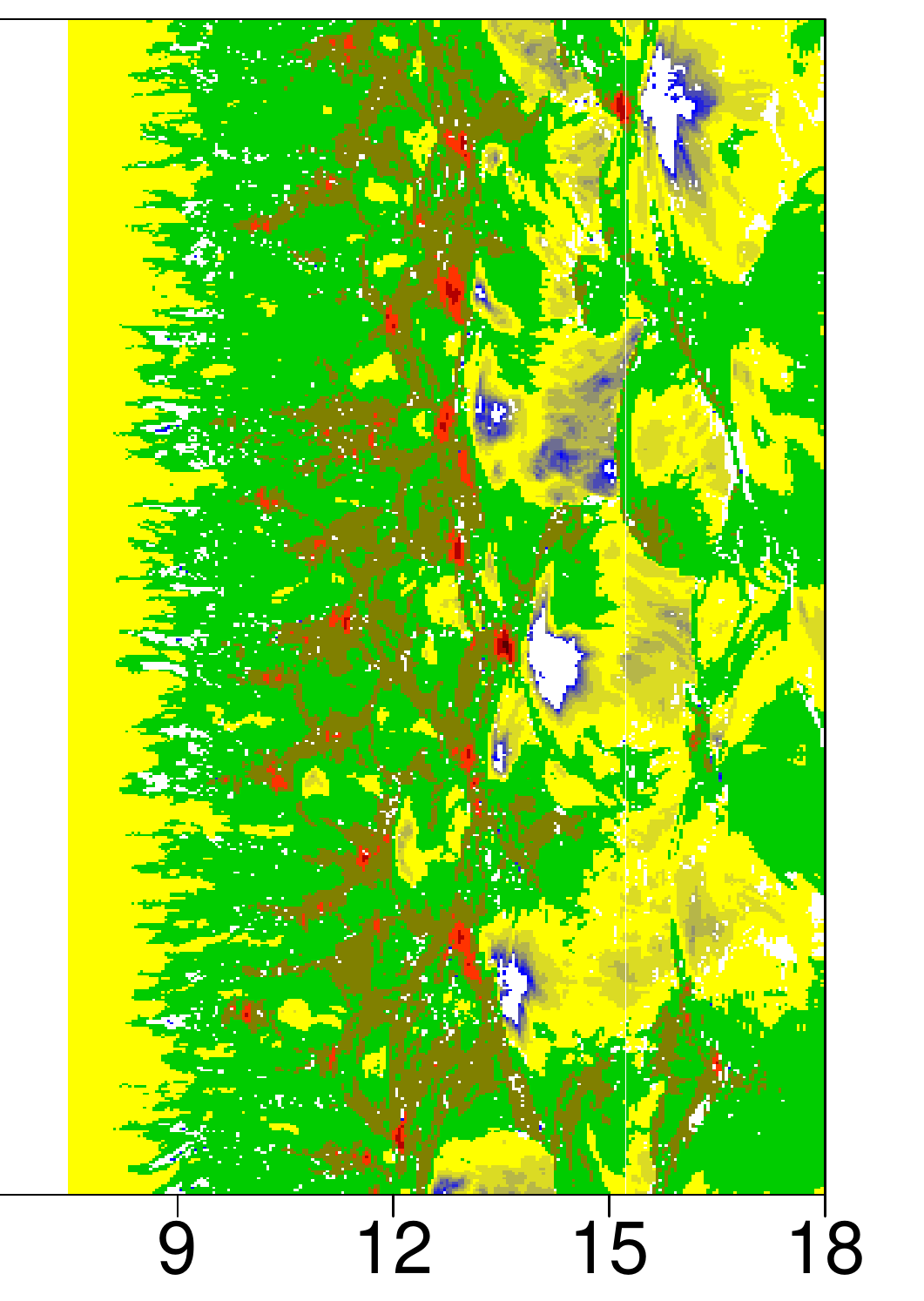}}
\put( 157,-30){\includegraphics[  height=5.5cm,trim= 15cm 1.3cm 0cm .4cm,clip]{{new_LD_timeseries_CAPE}.pdf}}

\put(11,-63){{\small \color{black} $t/h=13.2$}}
\put(43,-63){{\small \color{black} $13.6$}}
\put(69,-63){{\small \color{black} $13.9$}}

\put(-65,56){{\bf a}}
\put(-5,56){{\bf b}}
\put(40,56){{\bf c}}
\put(105,56){{\bf d}}

\put(-65,-53){{\bf e}}

\put(0,-53){{\bf f}}

\put(100,-53){{\bf g}}
\put(120,-58){(i)}
\put(120,-97){(ii)}
\put(120,-129){(iii)}

\end{overpic}
\vspace{280pt}
\caption{\small{\bf Dynamics of moisture convergence and potential energy.}
{\bf a}, Time series (units of model hours) of vertically integrated moisture convergence rates in $kg\,m^{-2}s^{-1}$ at altitudes below 2000 $m$ for CTR simulation, obtained by fixing one spatial coordinate. The remaining spatial as well as the time coordinate (units of 24h) are shown on the vertical and horizontal axes, respectively.
Vertical black line indicates onset of precipitation.
{\bf b,c}, Similar to (a) but for P2K and LD, respectively.
In (c), note the longer extent of the time axis as well as the black box, which specifies a case study region.
{\bf d}, CAPE in $J\;kg^{-1}$ corresponding to (c), i.e. for the LD simulation. Note the white regions showing depletion of CAPE.
{\bf e}, Probability density functions of moisture convergence for all three simulations. Colors chosen as in Fig.~\ref{fig:summary_diurnal_cycle}. (i) and (ii) represent the PDF before and after the onset of precipitation. 
Faint curves represent the PDF of divergence (i.e. negative values) mirrored along the vertical axis.
{\bf f}, Spatial plots of a 30x30 $km^2$ region for three points in time as labeled.
Top and bottom columns show precipitation intensity and moisture convergence, respectively. The color bar indicates precipitation intensity in $mm/h$.
{\bf g}, Time series of the point indicated by the black arrow in (c). (i) Convergence and precipitation intensity, (ii) CAPE, (iii) CIN.
}
\end{center}
\label{fig:local_convergence}
\end{figure}

This analysis highlights two aspects: 
First, precipitation intensity is not a simple function of surface temperature, and second, domain-average CAPE and CIN are not sufficient predictors of precipitation.
Consider therefore the space-time evolution of low-level moisture convergence (Fig.~\ref{fig:local_convergence}a---c), where one spatial coordinate was fixed.
The pattern can be described as a river-like network, with small streams of convergence merging to form larger, more intense streams, but finally dissipation of all structure.
We consider three stages of convective development:
(i) Before onset of precipitation (marked in Fig.~\ref{fig:local_convergence}a---c) moisture convergence fluctuates at small spatial scales and low magnitude.
The probability density function (PDF) is approximately symmetric around zero 
(Fig.~\ref{fig:local_convergence}e,top).
(ii) While precipitation occurs, the convergence pattern begins to organize, with areas of precipitation events serving as sources of divergence, and areas of convergence leading to the birth of new, and often larger, precipitation events. 
Notably, the PDF of convergence becomes skewed and acquires a shoulder of intense convergence, a feature not mirrored by divergence (Fig.~\ref{fig:local_convergence}e,bottom).
This is a signature of the asymmetry in the topological pattern (Fig.~\ref{fig:local_convergence}f), where convergence is characterized by thin filaments but divergence by patches.
(iii) After precipitation ceases, the convergence pattern again becomes featureless with low magnitude and a symmetric PDF (Fig.~\ref{fig:convergence_distribution}).

A distinguishing quality between the three simulations is that, in LD and P2K, convergence patterns have larger cross sections and are formed by more intense convergence and divergence (Figs~\ref{fig:local_convergence}a--c and \ref{fig:convergence_distribution}). 
Comparing CTR and LD, one might expect similar responses to the boundary conditions, since surface temperatures are similar. 
However, the simulated responses are quite different, indicating it is not the instantaneous boundary conditions, but the accumulation of energy in the system that drives a transition to a different regime.
These different responses are also reflected in the different behavior of the surface fluxes (Fig. \ref{fig:surfacefluxes}).
Comparing with the temporal evolution of precipitation intensity along the same coordinates as the moisture convergence (Fig.~\ref{fig:precip_intensity_vs_time}), we find that precipitation events --- originating near the location of strong convergence --- tend to be larger and more intense later in the day. 
Interestingly, although the prescribed diurnal surface temperature is very different for P2K and LD, the pattern produced is qualitatively quite similar. 

Cold pool currents have been suggested to facilitate event formation where they collide \citep{tompkins2001organizationCold,boing2012influence,schlemmer2014}.
Examining one particular area of the simulation domain for several times leading up to the birth of a precipitation event (Fig.~\ref{fig:local_convergence}f), the divergence from several previous events generates pronounced convergence in the vicinity.
Our simulations suggest that the generation of intense convergence is the mechanism that drives the observed, gradual ``coarse-graining'' of the spatial pattern, and thereby precipitation event size and intensity.

CAPE analyzed at grid point scale (in contrast to domain averages in Fig.~\ref{fig:summary_diurnal_cycle})
yields a strongly inhomogeneous spatial pattern, with values fluctuating by several orders of magnitude at any given time (Fig.~\ref{fig:local_convergence}d).
The fluctuations are strongest after onset of precipitation, and extremes of CAPE most pronounced when its average is already on the decline (Fig.~\ref{fig:CAPE_CDFS}).
Further, there are pronounced spatial correlations, with areas of depleted CAPE spreading out. 
Temporally, the pattern is anti-correlated with points of extreme CAPE 
often followed by near-complete removal of CAPE (consider a typical timeseries in Fig.~\ref{fig:local_convergence}g).
Our results suggest that interactions between convective precipitation events are critical in structuring the atmospheric moisture and energy distribution.

\begin{figure}[t]
\begin{center}
\vspace{1cm}
\begin{overpic}[width=5cm,angle=-90,trim= 0cm 0pt 0pt 0pt,clip]{dummy.pdf}
\put(-70,20){\includegraphics[height=3.9cm,angle=0,trim= 0cm 1.3cm 0cm 0cm,clip]{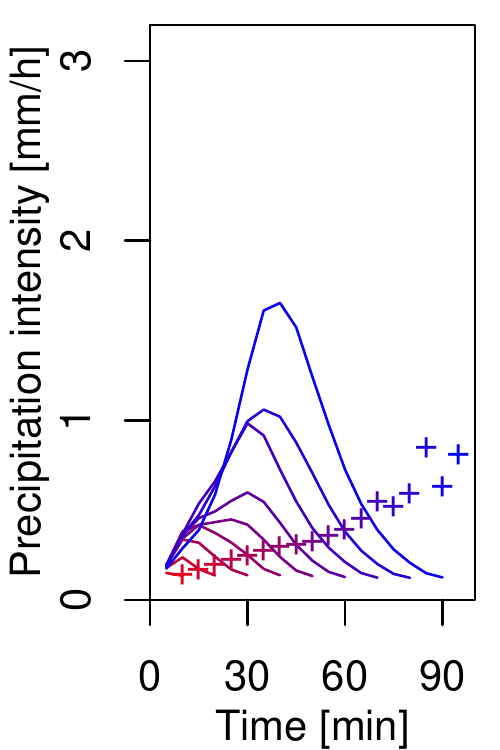}}
\put(-26.5,20){\includegraphics[height=3.9cm,angle=0,trim= 1.4cm 1.3cm 0cm 0cm,clip]{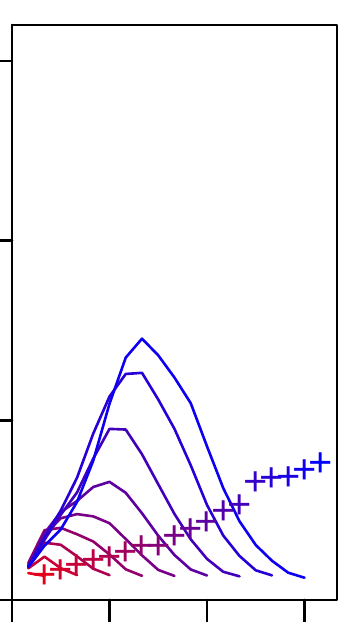}}
\put(5,20){\includegraphics[height=3.9cm,angle=0,trim= 1.4cm 1.3cm 0cm 0cm,clip]{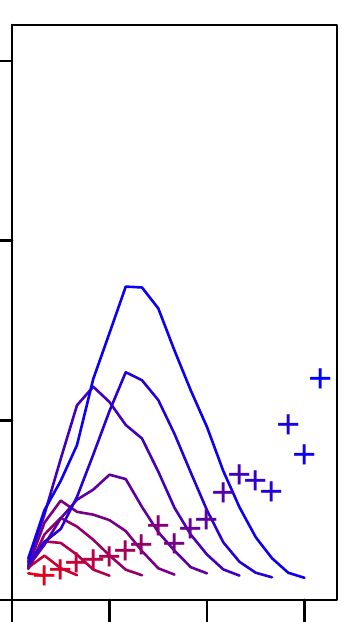}}

\put(-70,-54){\includegraphics[height=4.7cm,angle=0,trim= 0cm 0cm 0cm 0cm,clip]{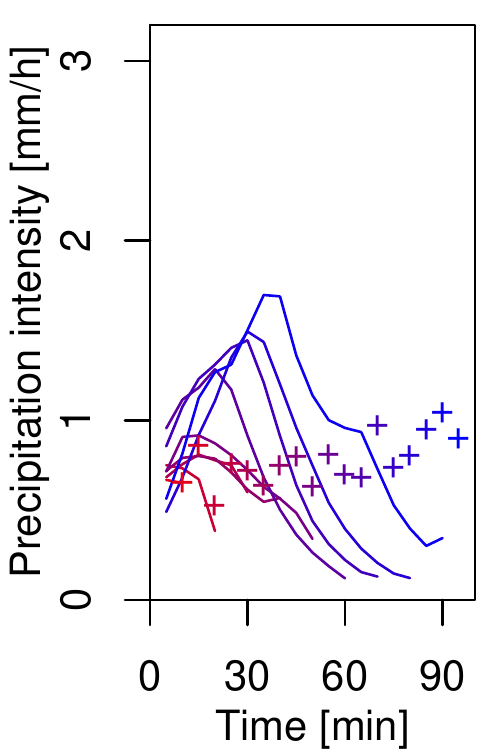}}
\put(-26.5,-54){\includegraphics[height=4.7cm,angle=0,trim= 1.4cm 0cm 0cm 0cm,clip]{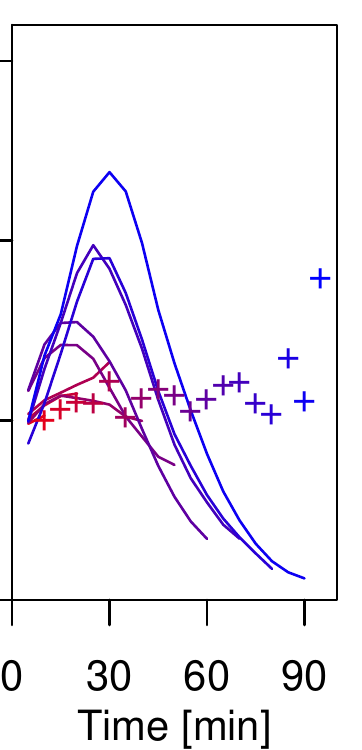}}
\put(5,-54){\includegraphics[height=4.7cm,angle=0,trim= 1.4cm 0cm 0cm 0cm,clip]{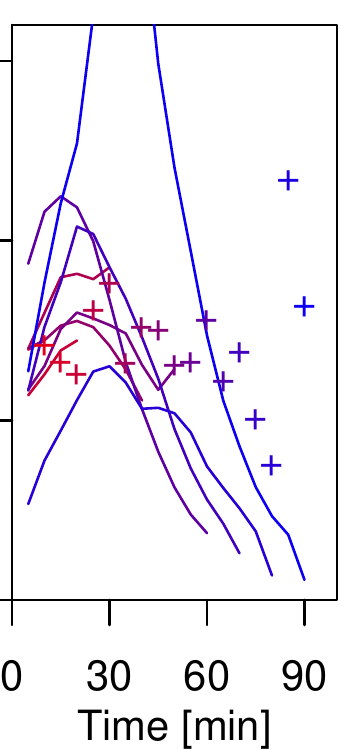}}


\put(-70,76){{\bf a}}
\put( -26,76){{\bf b}}
\put( 5,76){{\bf c}}

\put(-70,12){{\bf d}}
\put(-26,12){{\bf e}}
\put( 5,12){{\bf f}}

\end{overpic}
\vspace{120pt}
\caption{\small{\bf Surface rain intensity of tracks during their life cycles.} 
Average intensity of tracks for different durations. 
Shortest tracks (10 minutes) plotted as red curves, longest tracks (90 minutes) plotted as blue curves.
Each curve represents a ten-minute-interval of track durations.
Symbols (''+``) denote the average precipitation intensity for any track of corresponding color for five-minute binning of track data.
{\bf a}, CTR simulation for solitary tracks, i.e. tracks filtered such that they did not originate from, or terminate in, a collision of several tracks. 
{\bf b}, As (a) but for P2K.
{\bf c}, As (a) but for LD.
{\bf d}, Similar to (a) but for tracks originating from colliding tracks.
{\bf e}, Similar to (d), but for P2K. 
{\bf f}, Similar to (d), but for LD.
}
\end{center}
\label{fig:tracks_intensity}
\end{figure}

But how does this interaction impact on precipitation intensity?
Not all events are triggered by collisions of previous events, in particular not those forming during the initial onset of precipitation.  
We monitor interaction between cells of contiguous precipitation 
using a tracking algorithm (details: Methods), and record area and intensity of all precipitation event tracks for the three simulations. 
We distinguish events originating spontaneously and end without direct interactions with other events (termed ``solitary tracks''); and events emerging after collision of two or more events (termed ``mergers'').
For each track of a given type and life time we compute the average intensity within the duration of the track (Fig. \ref{fig:tracks_intensity}). 
Solitary tracks are very similar for all simulations (compare Fig.~\ref{fig:tracks_intensity}a---c), even though average precipitation intensity increased in LD and P2K relative to CTR (Fig.~\ref{fig:summary_diurnal_cycle}), i.e. those events that did not result from collisions are left approximately unaltered.
This is different for mergers: Track intensities are elevated already at initiation of the track and intensities are overall larger (Fig.~\ref{fig:tracks_intensity}d--f). 
Comparing the different simulations, intensities of mergers increase much more strongly in the case of P2K and LD as compared to CTR (see also Fig.~\ref{fig:rel_track_intensities}).
Mergers contribute most to the total precipitation intensity (averaged over the domain) for P2K and LD, while it is approximately balanced with the contribution for solitary tracks for CTR (Fig.~\ref{fig:prec_frac}). We note that some tracks are neither identified as solitary nor as mergers (i.e. tracks that split in their lifetime, or that merge into other cells). They provide a comparable amount of precipitation as the mergers.



Observational studies have recently ruled out simple thermodynamic measures, e.g. the Clausius-Clapeyron relation, as sufficient explanations for convective precipitation intensity increase with temperature \citep{Lenderink:2008,berg2013strong,molnar2015storm,westra2014future,wasko2015steeper}.
Our results imply that possible causes for convective invigoration cannot be sought in a model of a single convective plume \citep{westra2014future}, but a far more complex picture is required.
Even under unchanged average surface temperature dramatically elevated precipitation yield and intensity can be obtained, if the atmosphere is only given the time required for self-organization.
Characterizing the state of atmospheric self-organization at any given time hence requires the knowledge of past processes, i.e. the current state carries the memory of previous interactions. 
For classical proxies of convective potential, e.g. domain-averaged values of potential energy, our results imply that neither relatively low values of CAPE nor the presence of substantial CIN necessitate inhibition of intense precipitation.
Conversely, even when sufficient CAPE is present and triggering does occur, insufficient self-organization will prevent the atmosphere from harvesting all available CAPE.

Global climate models are typically based on mass flux parameterizations of convective processes, thereby relying on CAPE calculated at the coarse grid space. 
These models are hence incapable of simulating the sub-grid organization of convection \citep{mapes2011parameterizing}, and first parameterization efforts now incorporate cold pool dynamics into general circulation models \citep{rio2009shifting,grandpeix2010density}.
The impact of organization has been shown to be critical for tropical convection \citep{tan2015increases}, and likely has a significant effect on the Earth's albedo and radiative balance \citep{mauritsen2015missing}.
Convective organization is present on several scales, ranging from the interaction of neighboring convective updrafts at a scale of 1-10 km, over mesoscale convective systems (100-500 km), superclusters (1000-3000 km) up to the scale of the Madden-Julian Oscillation (MJO) with extents up to 10000 km \citep{moncrieff2013}. Our present study addresses rather the lower end of this cascade, and might be relevant for parametrizations on the scale of a typical GCM grid box.
Accurate modeling of the global radiation balance requires proper description of the diurnal cycle of convection, and self-organization has long been discussed as key to its understanding.
To incorporate the state of self-organization into lower-resolution models, e.g. global or regional climate models, we suggest that variables should be defined that capture the spatial pattern of crucial observables, e.g. moisture convergence or local CAPE and CIN, and the gradual increase of scales and interaction.

\section{Materials and methods}\label{sec:methods}
{\bf Large-eddy simulations (LES):}
We use the University of California, Los Angeles (UCLA) Large Eddy Simulation (LES) model \citep{stevens2005} to simulate idealized diurnal cycles over a land surface. 
The model employs a delta four-stream radiation scheme \citep{pincus2009monte} and the two-moment cloud microphysics scheme of \citep{seifert2006two}. 
Subgrid-scale turbulence is parameterized after Smagorinsky. 
In the horizontal, the size of the domain is $1024\times 1024$ grid cells for the simulations CTR and P2K, and $512\times 512$  for LD. The horizontal resolution is 200 $m$ and periodic boundary conditions are imposed. The lower boundary condition is homogeneous. 
In the vertical, 75 levels are used with a spacing of 100~$m$ below 1 $km$ stretching to 200 $m$ around 6 $km$ and 400 $m$ in the upper layers. 
The model top is located at 16.5 $km$ with a sponge layer above 12.3 $km$. 
For each simulation, the model is initialized by a homogeneous temperature and humidity profile idealized from radio soundings taken at the super site Lindenberg during the two summers of 2007-2008. 
Only potentially convective days, defined as days with a CAPE value larger than 1000 J kg$^{-1}$ at 12 Z are retained. 
The resulting averaged profiles of temperature and moisture at 0 Z are shown in Fig. \ref{fig:sounding}a and b together with the idealized profiles employed to start the simulation. 
The idealized temperature profile is obtained by prescribing a temperature lapse rate of 6.6 $K$ $km^{-1}$ below 11 $km$ and 3 $K$ $km^{-1}$ above 11 $km$ starting from a surface temperature of 21$^\circ$C. 
The profile of relative humidity is idealized as follows: 
Relative humidity linearly increases by 7.5$\% \, km^{-1}$ below 2 $km$, decreases by 12.5$\% \, km^{-1}$ between 2 and 4 $km$, by 2$\% \, km^{-1}$ between 4 and 10 $km$ and by 12$\% \, km^{-1}$ higher up. 
Surface relative humidity is set to 65\%. 
Knowing temperature and relative humidity, the idealized profile of specific humidity is obtained. 
Although more complicated, this procedure gives a better agreement to the observed specific humidity than directly specifying the specific humidity. 
Wind is set to zero.
In order to break translational symmetry, weak random noise is added to the initial profiles. 
At the surface, the sensible and latent heat fluxes are computed interactively by the model using Monin-Obukhov similarity theory and a prescribed skin temperature $T_{\rm skin}$. 
The diurnal cycle of $T_{\rm skin}$, measured at Lindenberg and averaged over the previously defined convective days, is approximated by the following relationship:
 \begin{eqnarray}
 T_{\rm skin}=T_o+10\sin((t-6)\pi/12)
 \end{eqnarray}
 with $t$ time in units of hours and $T_o$ the reference temperature (see Fig. \ref{fig:sounding}c). 
 The reference temperature amounts to 23 $^\circ$C in CTR. 
 To account for the fact that evaporation does not occur at its potential rate at Lindenberg, the saturation specific humidity is reduced by 30 percent. 
 The chosen values give a daily averaged sensible and latent heat flux of 21.8 and 67.1 W m$^{-2}$ in CTR versus 13.8 and 71.9 W m$^{-2}$ at Lindenberg. 
 As we start from no background wind, the surface fluxes are zero in UCLA during night, whereas the observed latent heat flux amounts to 30 W m$^{-2}$.  
 During daytime, both fluxes are overestimated in CTR as compared to the observations, but the Bowen ratio remains similar. 
 The Bowen ratio amounts to 0.33 in CTR and 0.32 in the observations at 12 LST.
 
All simulations are initialized at 0 UTC and run over one diurnal cycle, which is triggered by solar insolation at the latitude of Lindenberg (52$^\circ$N). 
We performed the following simulations: $T_o=23$ (CTR), $T_o=24$ (P1K), $T_o=25$ (P2K), as well as a simulation with identical boundary conditions to CTR but an overall stretch of the model day by a factor of two.
As a sensitivity study, we further repeated CTR with a smaller domain of $1/2$ the linear dimension, i.e. 512$\times$512 grid boxes.

The 3D fields of prognostic variables are output instantaneously in a 5 minute interval.

{\bf Rain cell tracking}:
For our statistical analysis we view rain events as disparate spatial entities in space, i.e. they are isolated from one-another by rainless spaces. 
To distinguish them, a tracking algorithm is required that can identify contiguous rain patches, and link patches between two consecutive output time steps when they belong to the same rain event. 
Here, we apply an improved version of the ``iterative raincell tracking`` (IRT), originally developed for tracking of rain events as seen from a weather radar. 
Since mean wind is absent in our simulation setup, the tracking is simplified in the sense that we do not have to take into account that patches are displaced by the dominant advection field between consecutive time steps. 
Therefore, in contrast to an application on radar data, the tracking does not have to be iterated multiple times, and one run is sufficient (details: \citep{moseley2014}).
In the optimized version of IRT used here, a shortcoming in the treatment of merging and splitting events has been improved: 
After a splitting event occurs, the largest fragment continues the original track, if the area of the second largest fragment is smaller or equal than $c$ times the area of the largest fragment. All other fragments begin new tracks.
Analogously, the largest track that is participating in a merging event is continued by the merged track, if the second largest participant is smaller or equal than $c$ times the area of the largest one. All other participants of the merging event terminate. Here, we use $c=1/4$ (the old version of the code is equivalent to $c=0$).
We apply the tracking to the rainfall intensity at the surface with a threshold of 0.1 $mm\, h^{-1}$.


{\bf CAPE analysis}:
We estimate the {\em Convective Available Potential Energy} (CAPE) and the {\em Convective Inhibition} (CIN) from the profiles (domain mean for results in Fig. \ref{fig:summary_diurnal_cycle}, column wise for results in Fig. \ref{fig:local_convergence}) of temperature, water vapor mixing ratio, cloud water mixing ratio, and pressure. An imaginary air parcel is adiabatically lifted from the lowest model level to the model top. Below the {\em Level of Free Convection} (LFC), the virtual potential temperature of the air parcel $\theta_{v,\mathrm{parcel}}$ is lower than that of the environment which we here approximate by the domain mean $\theta_v$, meaning that it is negatively buoyant. Above the LFC, where $\theta_{v,\mathrm{parcel}}>\theta_v$, it positively buoyant, up to a height where $\theta_{v,\mathrm{parcel}}<\theta_v$ again, which we call the {\em Limit Of Convection} (LOC). The virtual potential temperature is defined by
\begin{equation}
\theta_v=\theta\, \left[1+\left(\frac{R_v}{R_d}-1\right)r-r_l\right]
\end{equation}
In the above equation, $r$ and $r_l$ are the mixing ratios of water vapor and cloud water, and $R_d$ and $R_v$ are the ideal gas constants of dry air and water vapor, respectively. The potential temperature is defined by
\begin{equation}
\theta=T\left(\frac{p_0}{p}\right)^{\frac{R_d}{c_p}} \; ,
\end{equation}
where $T$ is the temperature, $p$ is the pressure, $p_0=1000\,hPa$ is a reference pressure, and $c_p$ is the specific heat capacity at constant pressure of dry air. CIN is integrated from the surface to the LFC where the parcel is negatively buoyant:
\begin{equation}
\mathrm{CIN}=\int_0^{z_\mathrm{LFC}}\frac g\theta (\theta_v-\theta_{v,\mathrm{parcel}})dz
\end{equation}
CAPE, on the other hand, is integrated from LFC up to LOC, where the parcel is positively buoyant:
\begin{equation}
\mathrm{CAPE}=\int_{z_\mathrm{LFC}}^{z_\mathrm{LOC}}\frac g\theta (\theta_{v,\mathrm{parcel}}-\theta_v)dz \; .
\end{equation}

While the air parcel is lifted, its enthalpy and its total water content is kept constant. Below the {\em Lifted Condensation Level} (LCL) the parcel is unsaturated, and its potential temperature $\theta_\mathrm{parcel}$ is approximately constant. At LCL the air is saturated, and remains saturated as the parcel is lifted further above while the excessive water vapor is condensed to cloud water under latent heat release. From there upward the parcels follows a {\it pseudo-adiabatic} lapse rate.
Further details are given in \citep{stull}.

\renewcommand{\thesection}{S\arabic{section}}

\renewcommand{\thefigure}{S\arabic{figure}}
\setcounter{figure}{0}

\renewcommand{\thetable}{S\arabic{table}}
\setcounter{table}{0}

\renewcommand{\theequation}{S\arabic{equation}}
\setcounter{table}{0}

\setcounter{page}{1}

\begin{figure}[h]
\begin{center}
\begin{overpic}[width=5cm,angle=-90,trim= 0cm 0pt 0pt 0pt,clip]{dummy.pdf}
\put(-70,-150){\includegraphics[height=20.0cm,angle=0,trim= 0cm 0cm 0cm 0cm]{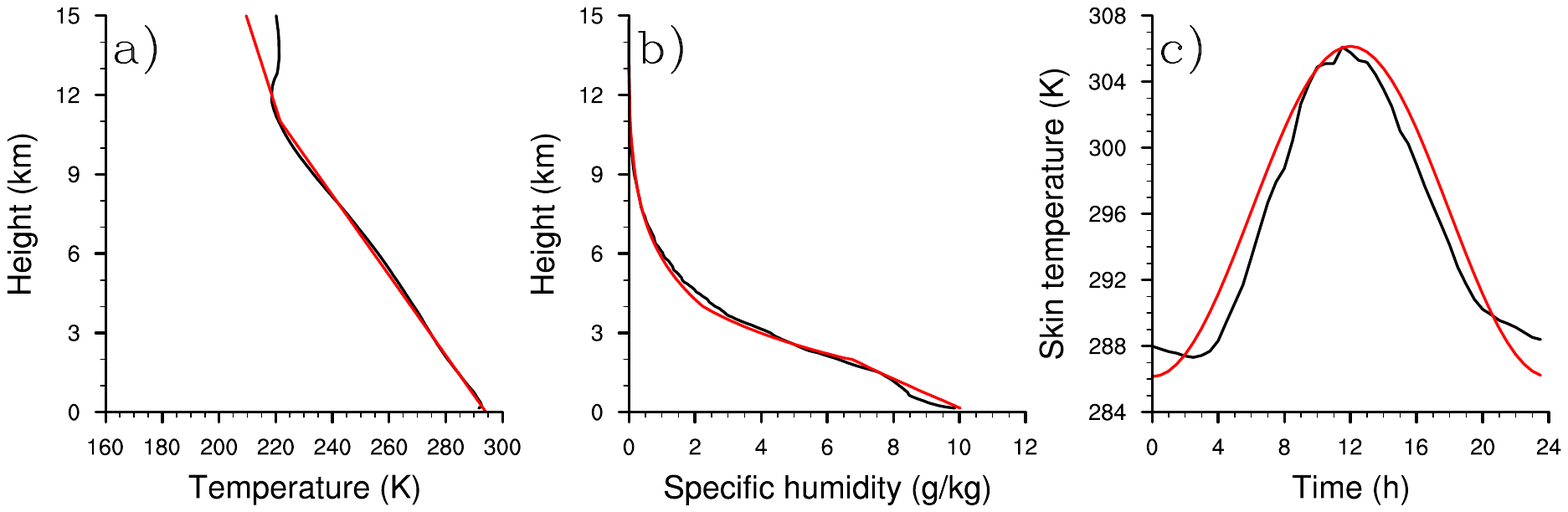}}
\end{overpic}
\vspace{150pt}
\caption{{\bf Vertical profiles.} 
{\bf a,b}, averaged profiles of temperature and moisture at 0 Z (black) and idealized profiles (red). 
{\bf c}, Diurnal cycle of skin temperature.}
\label{fig:sounding}
\end{center}
\end{figure}

\begin{figure}[h]
\begin{center}
\begin{overpic}[width=5cm,angle=-90,trim= 0cm 0pt 0pt 0pt,clip]{dummy.pdf}
\put(-70,-70){\includegraphics[height=6.0cm,angle=0,trim= 0cm 0cm 0cm 0cm]{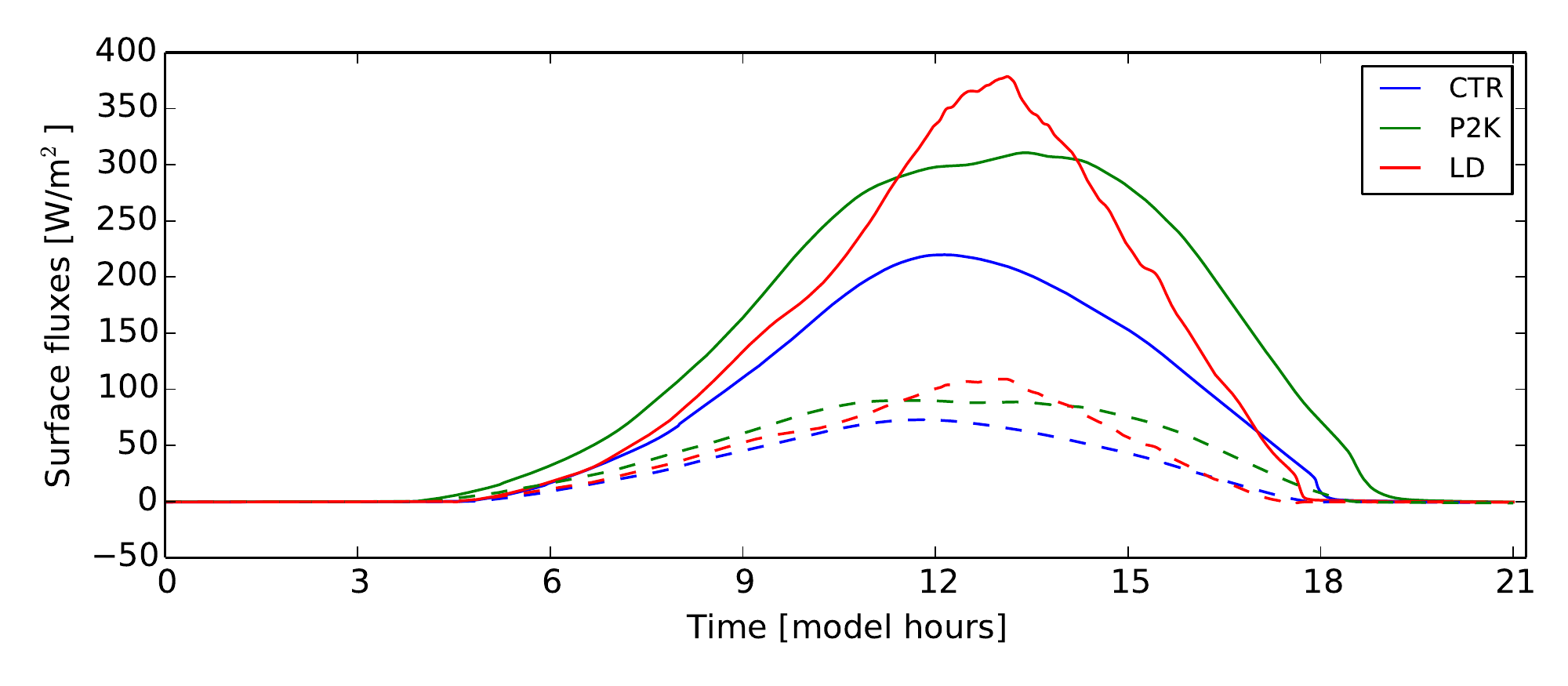}}
\end{overpic}
\vspace{150pt}
\caption{{\bf Surface fluxes.} 
Surface latent (solid) and sensible (dashed) heat fluxes, averaged over the entire domain, for the CTR, P2K, and LD simulations.}
\label{fig:surfacefluxes}
\end{center}
\end{figure}

\begin{figure}[t]
\begin{center}
\begin{overpic}[width=5cm,angle=-90,trim= 0cm 0pt 0pt 0pt,clip]{dummy.pdf}

\put( -70,-80){\includegraphics[height=9cm,trim= 0cm 0pt 8cm 1.5cm,clip]{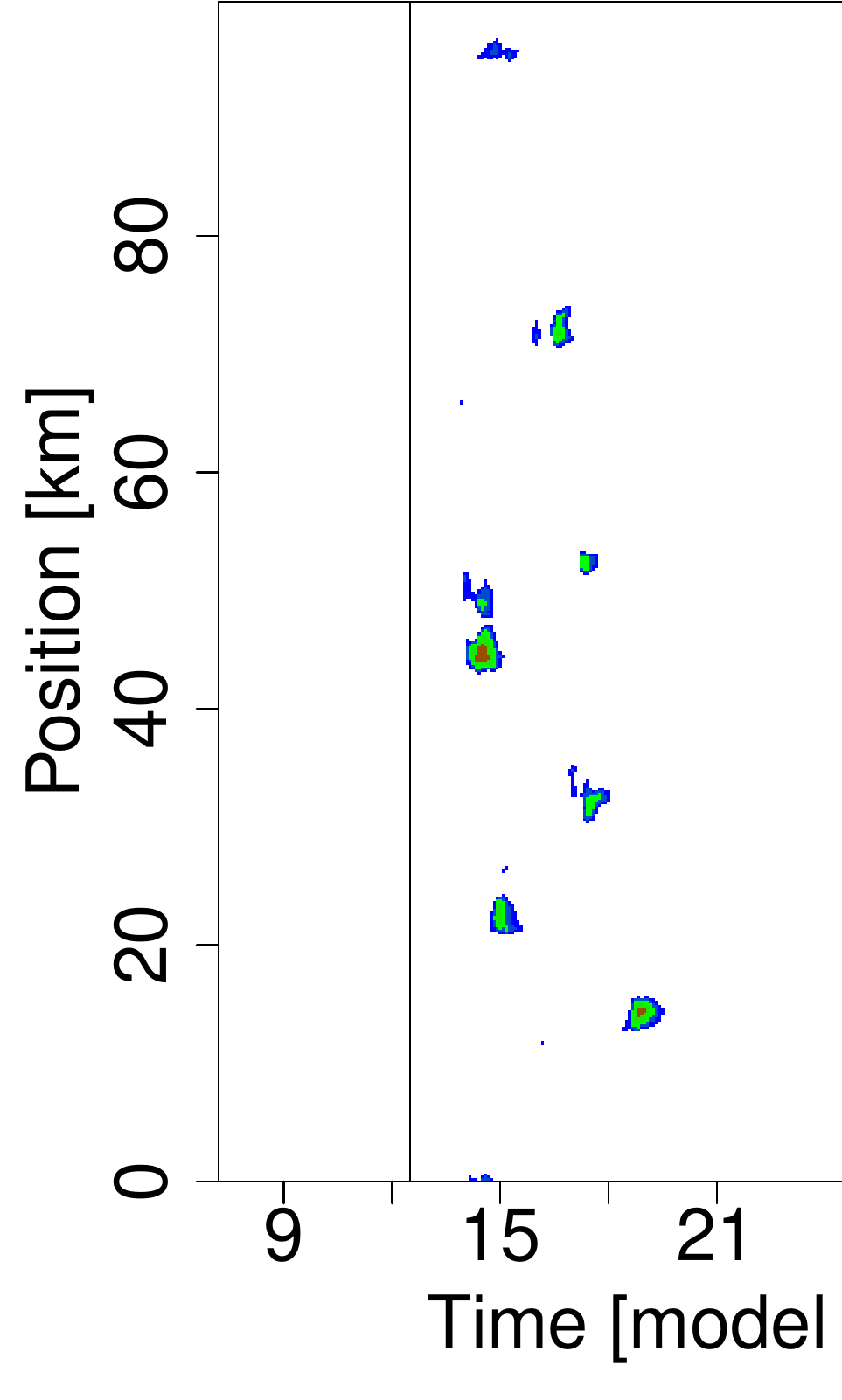}}
\put(   10,-80){\includegraphics[height=9cm,trim= 2.5cm 0pt 8cm 1.5cm,clip]{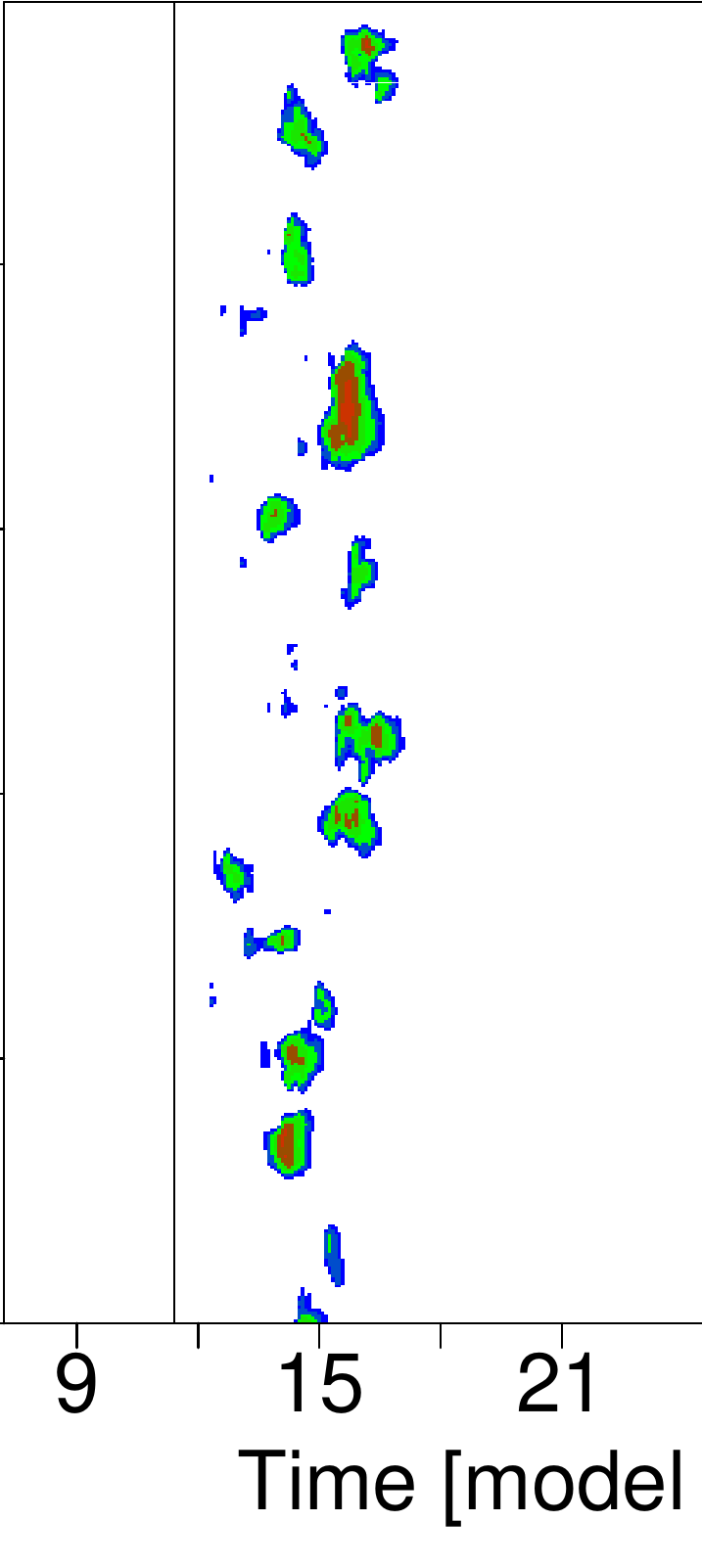}}
\put(  70,-80){\includegraphics[height=9cm,trim= 5cm 0pt 0cm 1.5cm,clip]{{LD_r_int_timeseries_300_1_500}.pdf}}

\put(-73,50){{\bf a}}
\put( 10,50){{\bf b}}
\put(77,50){{\bf c}}

\end{overpic}
\vspace{170pt}
\caption{{\bf Precipitation intensity vs. time.}
Units of $mm\;h^{-1}$.
{\bf a}, CTR simulation.
{\bf b}, P2K simulation.
{\bf c}, LD simulation.
}
\end{center}
\label{fig:precip_intensity_vs_time}
\end{figure}

\begin{figure}[t]
\begin{center}
\begin{overpic}[width=5cm,angle=-90,trim= 0cm 0pt 0pt 0pt,clip]{dummy.pdf}
\put(-17,-46){\includegraphics[height=9.8cm,trim= 0cm 0pt 0cm 0cm,clip]{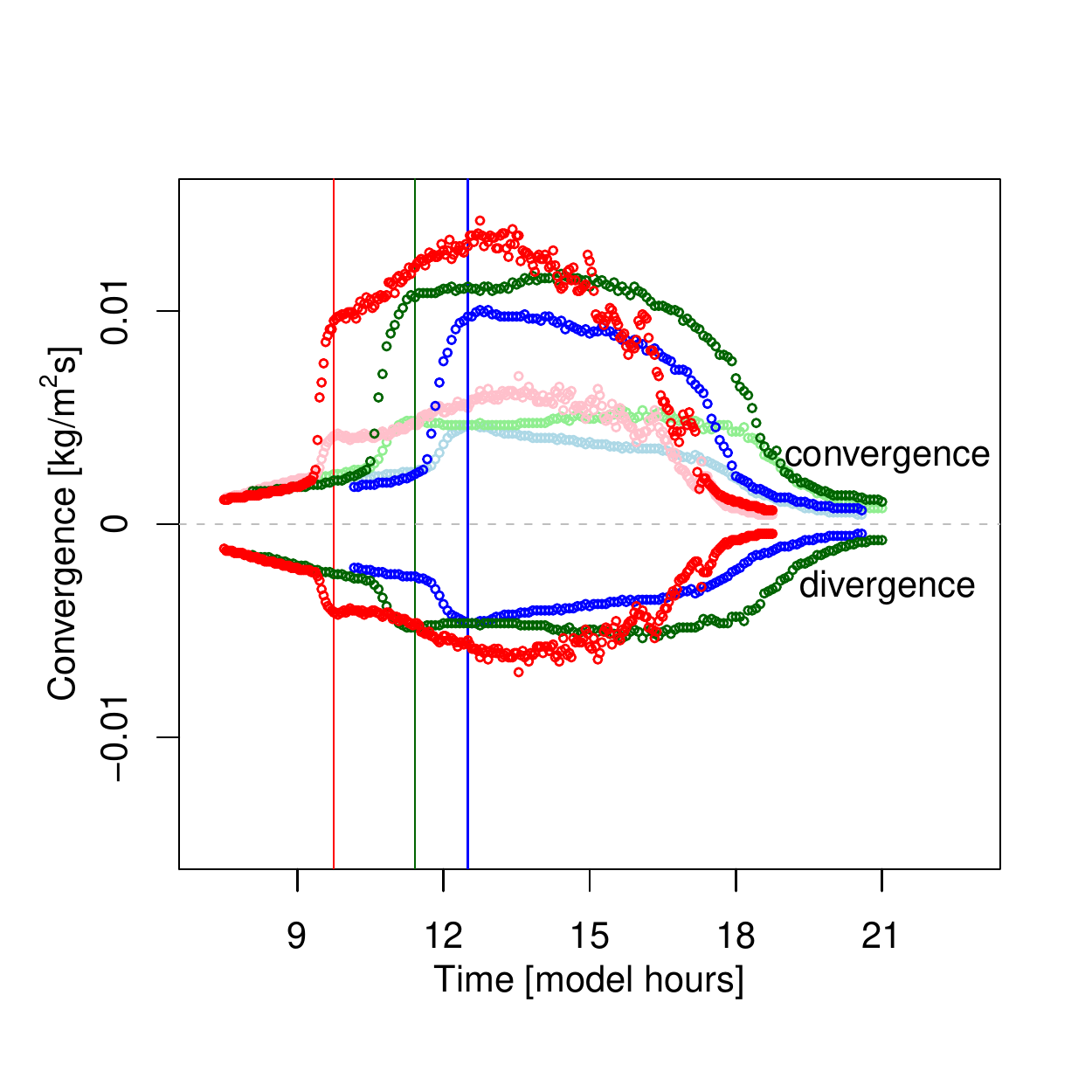}}

\end{overpic}
\vspace{60pt}
\caption{{\bf Diurnal cycle of moisture convergence and divergence.} 
Extremes of convergence (positive values) and divergence (negative values) and mirrored values (faint).
Extremes defined here as the 99.9th percentiles of the convergence and divergence distribution, respectively.
Blue, green, and red colors mark the different simulations CTR, P2K, and LD, respectively.
Vertical lines mark the onset of precipitation for the simulations of corresponding colors.
}
\end{center}
\label{fig:convergence_distribution}
\end{figure}

\begin{figure}[t]
\begin{center}
\vspace{1cm}
\begin{overpic}[width=5cm,angle=-90,trim= 0cm 0pt 0pt 0pt,clip]{dummy.pdf}
\put(-70,20){\includegraphics[height=3.9cm,angle=0,trim= 0cm 1.3cm 0cm 0cm,clip]{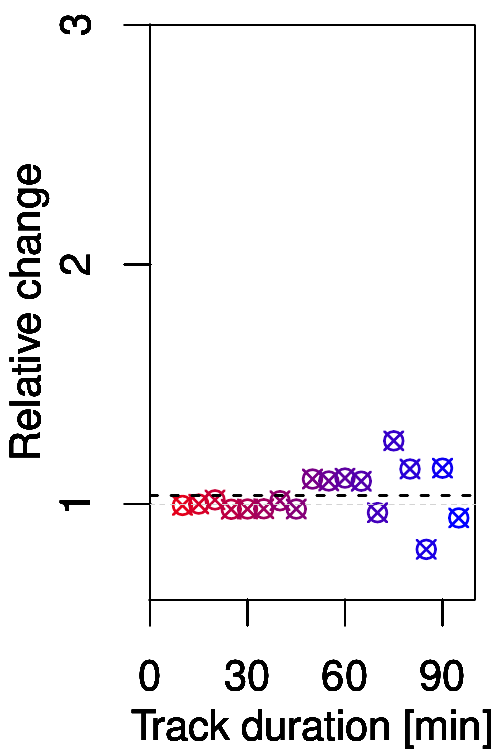}}
\put(-23.5,20){\includegraphics[height=3.9cm,angle=0,trim= 1.4cm 1.3cm 0cm 0cm,clip]{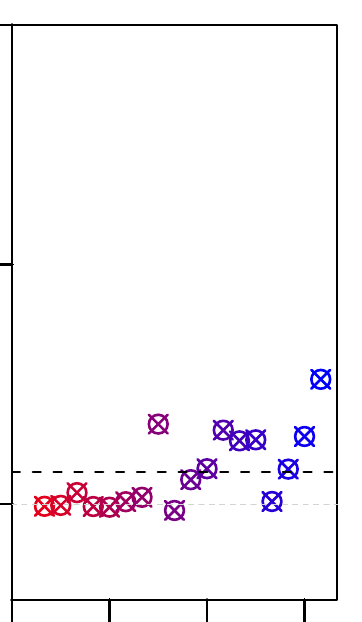}}

\put(-70,-54){\includegraphics[height=4.7cm,angle=0,trim= 0cm 0cm 0cm 0cm,clip]{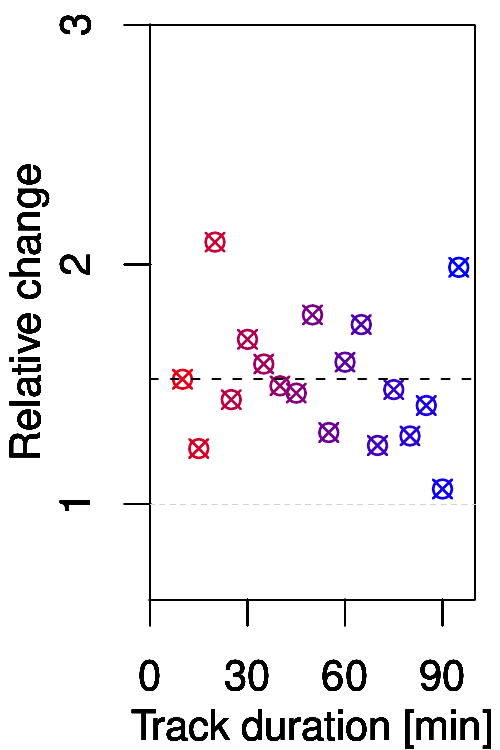}}
\put(-23.5,-54){\includegraphics[height=4.7cm,angle=0,trim= 1.4cm 0cm 0cm 0cm,clip]{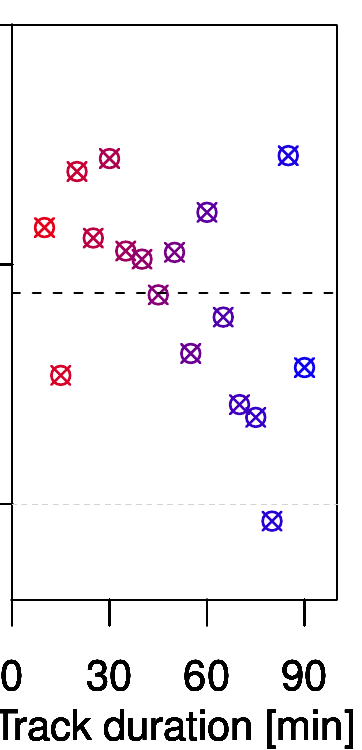}}

\put(-70,76){{\bf a}}
\put( -26,76){{\bf b}}

\put(-70,12){{\bf c}}
\put(-26,12){{\bf d}}

\end{overpic}
\vspace{120pt}
\caption{{\bf Relative change in surface rain intensity of tracks for different durations of tracks.} 
Ratios of average intensities of tracks for different durations, colors ranging from red to blue for short (10 min) to long (90 min) tracks. 
Each symbol represents a ratio of two corresponding averages.
{\bf a}, P2K simulation vs. CTR simulation for solitary tracks, i.e. tracks filtered such that they did not originate from, or terminate in, a collision of several tracks. 
{\bf b}, As (a) but for LD.
{\bf c}, Similar to (a) but for tracks originating from colliding tracks.
{\bf d}, Similar to (c), but for LD. 
In all panels, the dashed black lines indicate the average change for all durations.
The light gray lines indicate unity, i.e. a unchanged averages, as a guide to the eye.
}
\end{center}
\label{fig:rel_track_intensities}
\end{figure}

\begin{figure}[h]
\begin{center}
\begin{overpic}[width=5cm,angle=-90,trim= 0cm 0pt 0pt 0pt,clip]{dummy.pdf}
\put(-70,-70){\includegraphics[height=9.0cm,angle=0,trim= 0cm 0cm 0cm 0cm]{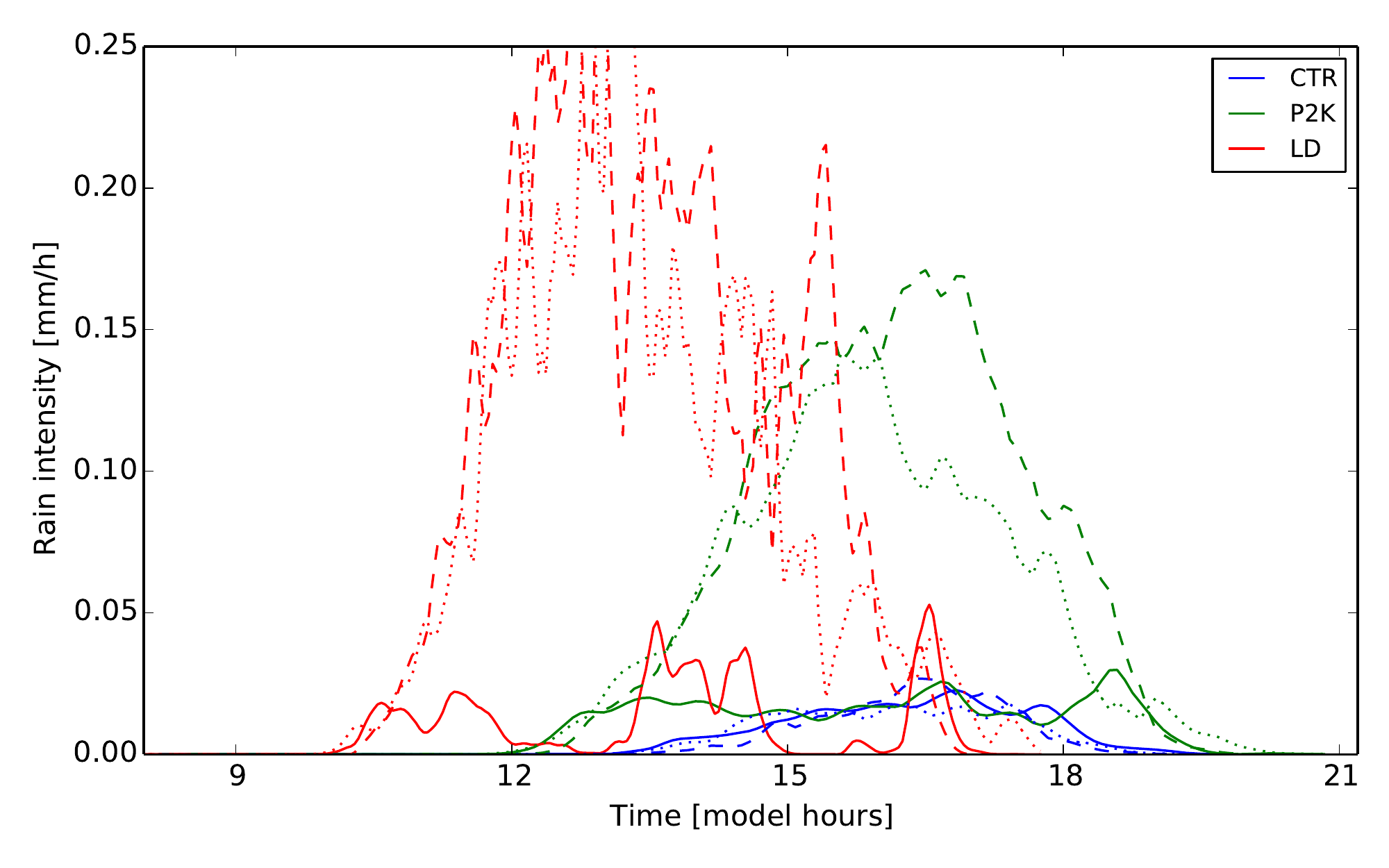}}
\end{overpic}
\vspace{150pt}
\caption{{\bf Contribution of track types to the total amount of precipitation.} Solitary tracks (solid), mergers (dashed), events which are neither identified as solitary nor as merger (dotted), for CTR, P2K, and LD simulation. }
\label{fig:prec_frac}
\end{center}
\end{figure}





\begin{figure}[t]
\begin{center}
\begin{overpic}[width=5cm,angle=-90,trim= 0cm 0pt 0pt 0pt,clip]{dummy.pdf}

\put( -70,-10){\includegraphics[height=7cm,trim= 0cm 0cm 0cm 0cm,clip]{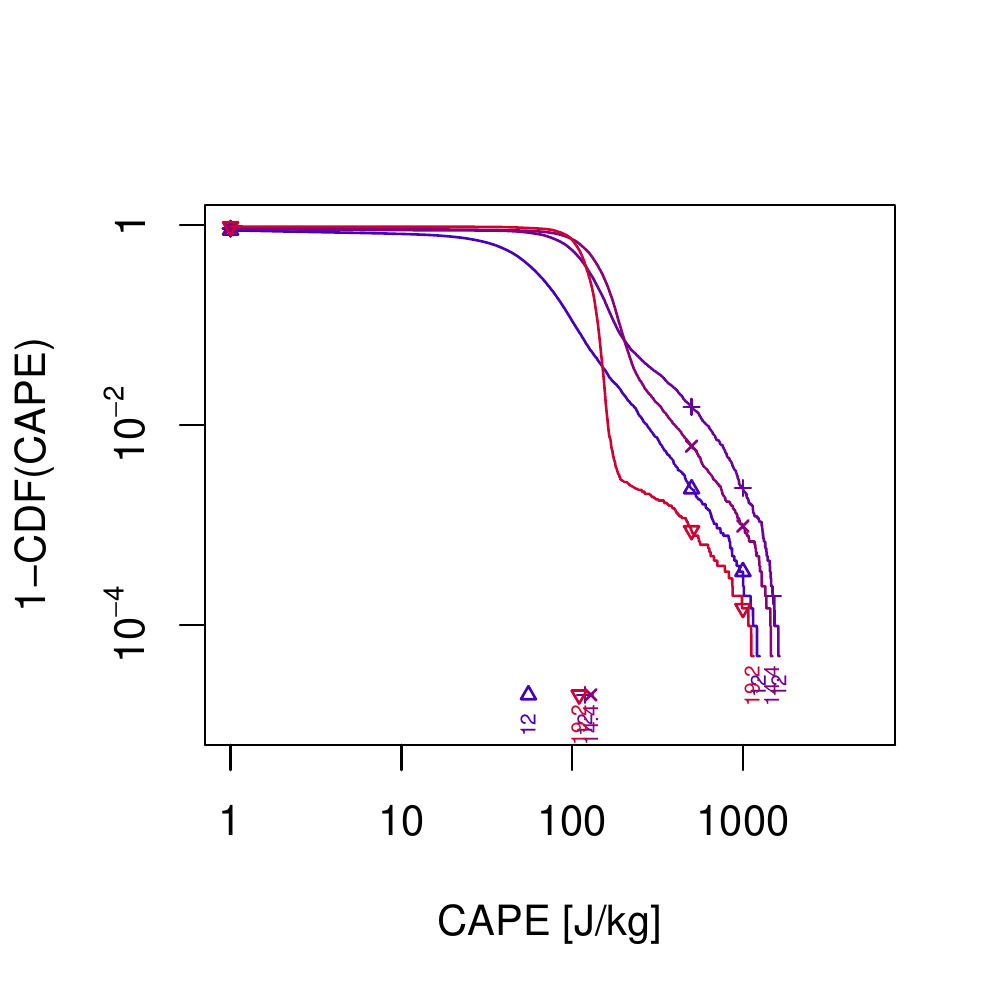}}
\put(  20,-10){\includegraphics[height=7cm,trim= 2cm 0cm 0cm 0cm,clip]{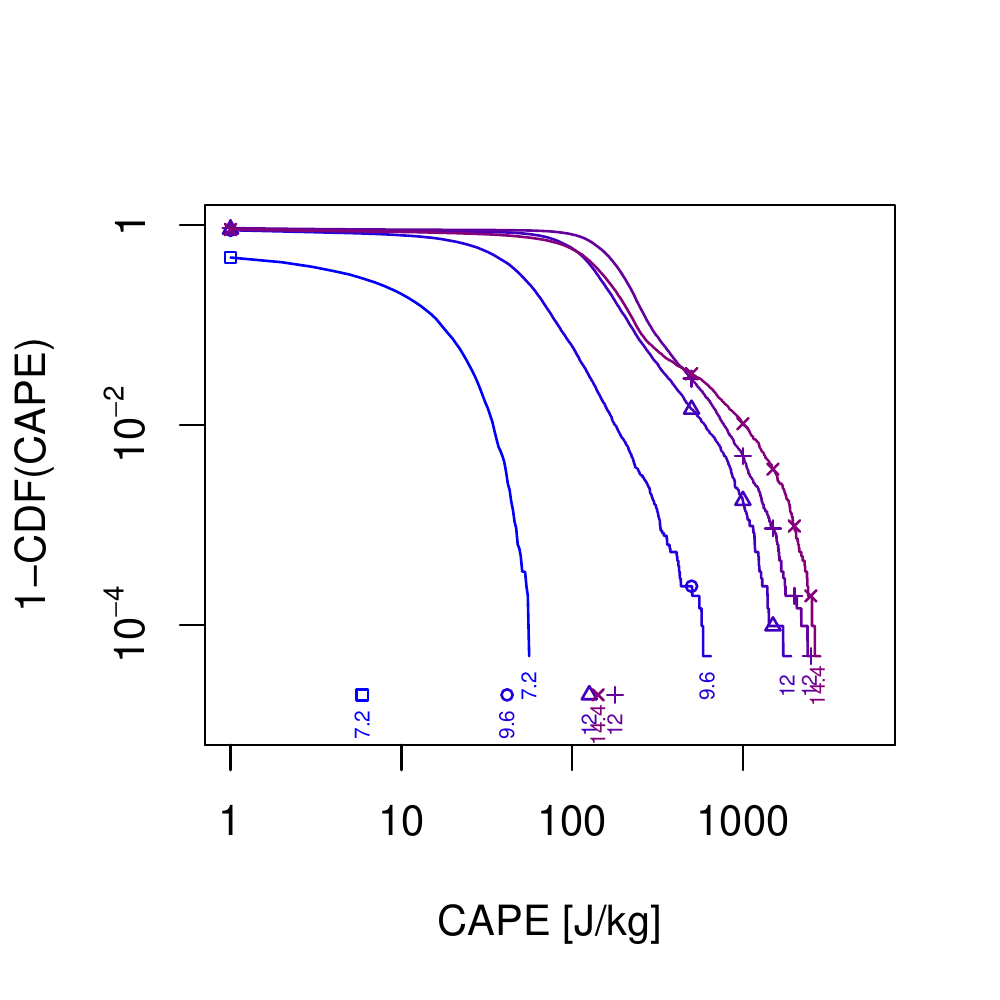}}
\put( 90,-10){\includegraphics[height=7cm,trim= 2cm 0cm 0cm 0cm,clip]{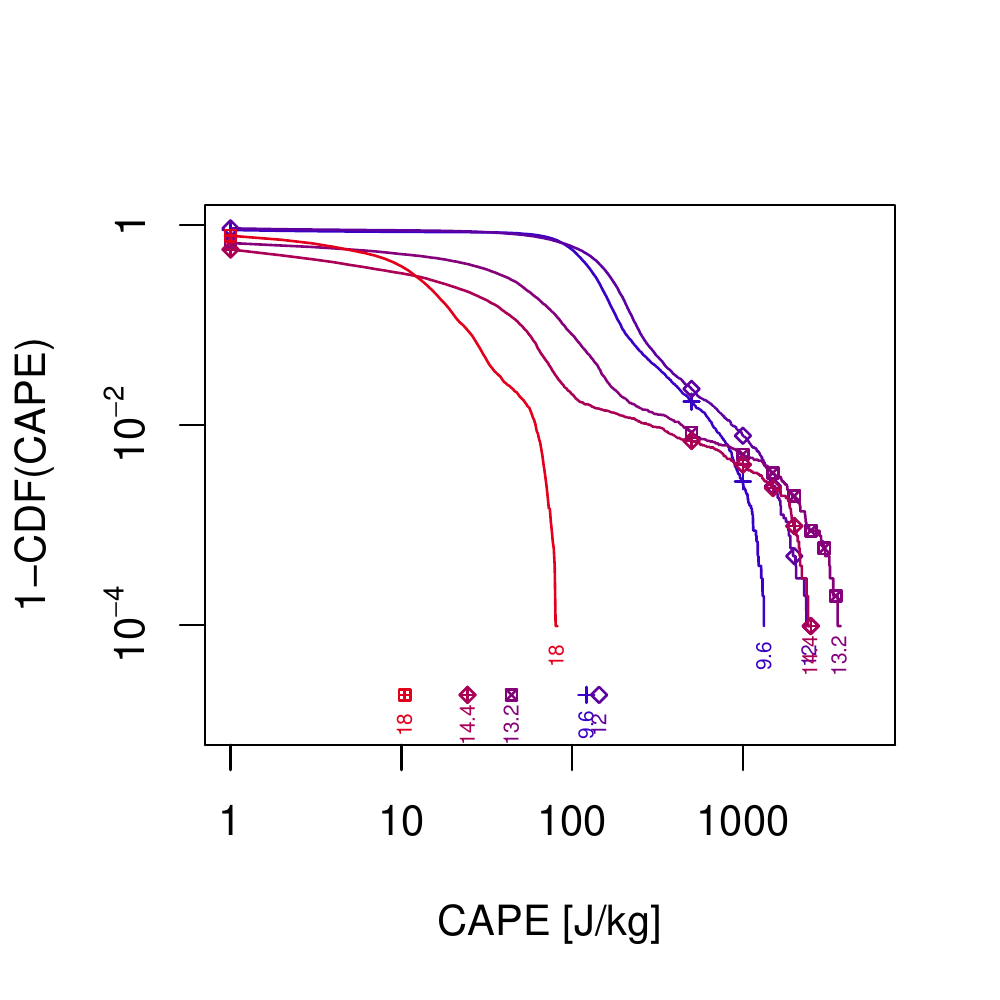}}
\put(-65,70){{\bf a}}
\put( 22,70){{\bf b}}
\put( 92,70){{\bf c}}
\end{overpic}
\vspace{20pt}
\caption{{\bf Probability density function of CAPE.}
{\bf a}, CTR simulation.
{\bf b}, P2K simulation.
{\bf c}, LD simulation.
Numbers indicate the time in units hours, where in the case of LD units were rescaled by a factor of 1/2 (as in Fig.~\ref{fig:summary_diurnal_cycle}).
Times chosen near the onset of precipitation, near the peak of average CAPE, as well as two later times (compare Fig.~\ref{fig:summary_diurnal_cycle}).
}
\end{center}
\label{fig:CAPE_CDFS}
\end{figure}


\end{article}
\bibliography{references}
\bibliographystyle{plainnat}
\end{document}